\documentclass[aps,twocolumn,prl,reprint,amsmath,amssymb,superscriptaddress]{revtex4-1}

\usepackage{graphicx}
\usepackage{dcolumn}
\usepackage{multirow}
\usepackage{hhline}
\usepackage{booktabs}
\usepackage{bm}
\usepackage{microtype}
\DeclareGraphicsExtensions{.pdf,.png,.jpg}

\begin{document}

\title{Tailoring supercurrent confinement in graphene bilayer weak links}

\author{Rainer Kraft}
\affiliation{Institute of Nanotechnology, Karlsruhe Institute of Technology, D-76021 Karlsruhe, Germany}

\author{Jens Mohrmann}
\affiliation{Institute of Nanotechnology, Karlsruhe Institute of Technology, D-76021 Karlsruhe, Germany}

\author{Renjun Du}
\affiliation{Institute of Nanotechnology, Karlsruhe Institute of Technology, D-76021 Karlsruhe, Germany}

\author{Pranauv Balaji Selvasundaram}
\affiliation{Institute of Nanotechnology, Karlsruhe Institute of Technology, D-76021 Karlsruhe, Germany}
\affiliation{Department of Materials and Earth Sciences, Technical University Darmstadt, Darmstadt, Germany}

\author{Muhammad Irfan}
\affiliation{Kavli  Institute  of  Nanoscience,  Delft  University  of  Technology,
P.O.  Box  4056,  2600  GA  Delft,  The  Netherlands}

\author{Umut Nefta Kanilmaz}
\affiliation{Institute of Nanotechnology, Karlsruhe Institute of Technology, D-76021 Karlsruhe, Germany}
\affiliation{Institute for Condensed Matter Theory, Karlsruhe Institute of Technology, D-76128 Karlsruhe, Germany}

\author{Fan Wu}
\affiliation{Institute of Nanotechnology, Karlsruhe Institute of Technology, D-76021 Karlsruhe, Germany}
\affiliation{College of Optoelectronic Science and Engineering, National University
of Defense Technology, Changsha 410073, China}

\author{Detlef Beckmann}
\affiliation{Institute of Nanotechnology, Karlsruhe Institute of Technology, D-76021 Karlsruhe, Germany}

\author{Hilbert von L\"{o}hneysen}
\affiliation{Institute of Nanotechnology, Karlsruhe Institute of Technology, D-76021 Karlsruhe, Germany}
\affiliation{Institute of Physics, Karlsruhe Institute of Technology, D-76049 Karlsruhe, Germany}
\affiliation{Institute for Solid State Physics, Karlsruhe Institute of Technology, D-76021 Karlsruhe, Germany}


\author{Ralph Krupke}
\affiliation{Institute of Nanotechnology, Karlsruhe Institute of Technology, D-76021 Karlsruhe, Germany}
\affiliation{Department of Materials and Earth Sciences, Technical University Darmstadt, Darmstadt, Germany}

\author{Anton Akhmerov}
\affiliation{Kavli  Institute  of  Nanoscience,  Delft  University  of  Technology,
P.O.  Box  4056,  2600  GA  Delft,  The  Netherlands}

\author{Igor Gornyi}
\affiliation{Institute of Nanotechnology, Karlsruhe Institute of Technology, D-76021 Karlsruhe, Germany}
\affiliation{Institute for Condensed Matter Theory, Karlsruhe Institute of Technology, D-76128 Karlsruhe, Germany}
\affiliation{A.F. Ioffe Physico-Technical Institute, 194021 St. Petersburg, Russia}

\author{Romain Danneau}
\affiliation{Institute of Nanotechnology, Karlsruhe Institute of Technology, D-76021 Karlsruhe, Germany}

\maketitle

\onecolumngrid

\textbf{The Josephson effect is one of the most studied macroscopic quantum phenomena in condensed matter physics and has been an essential part of the quantum technologies development over the last decades. It is already used in many applications such as magnetometry, metrology, quantum computing, detectors or electronic refrigeration. However, developing devices in which the induced superconductivity can be monitored, both spatially and in its magnitude, remains a serious challenge. In this work, we have used local gates to control confinement, amplitude and density profile of the supercurrent induced in one-dimensional nanoscale constrictions, defined in bilayer graphene-hexagonal boron nitride van der Waals heterostructures. The combination of resistance gate maps, out-of-equilibrium transport, magnetic interferometry measurements, analytical and numerical modelling enables us to explore highly tunable superconducting weak links. Our study opens the path way to design more complex superconducting circuits based on this principle such as electronic interferometers or transition-edge sensors.}



\hspace{0.5cm}

\twocolumngrid

Superconductivity can be induced in a material by direct contact to a superconductor. This proximity effect allows the transmission of Andreev pairs from a superconducting electrode to another when these are close enough. The Josephson effect can then be measured as it is observed in tunnel junctions \cite{Josephson1962,Anderson1963,tinkhambook}. However, the tuning of the dissipationless current in such Josephson junctions is not possible without changing its geometry or temperature. By replacing the tunnel junction by a so-called weak link \cite{Likharev1979,BaronePaternoBook}, \textit{i.e.} any kind of conductive system, the supercurrent may flow over a much larger distance than the couple of nanometers of a tunnel barrier. The magnitude of the supercurrent mainly depends on the contact transparency, the disorder in the weak link and the temperature \cite{Likharev1979}.

Many different types of materials and systems have been used as weak links, ranging from mesoscopic diffusive metallic wires \cite{Baselmans1999}, two-dimensional (2D) electron gas \cite{Schaepers2003}, graphene \cite{Heersche2007}, topological insulators \cite{Zhang2011,Sacepe2011,Veldhorst2012,Oostinga2013,Hart2014,Bocquillon2016} and quantum dots \cite{DeFranceschi2010}, as well as atomic contacts \cite{Goffman2000}. When graphene is utilised as a weak link, the Josephson effect can be tuned by electrostatic gating \cite{Heersche2007,Du2008,Miao2009,Rickhaus2012,Coskun2012,Komatsu2012,
Mizuno2013,Choi2013} and, thanks to edge connection which provides very low contact resistance \cite{Wang2013}, it is possible to measure large supercurrent as well as ballistic interferences \cite{Calado2015,BenShalom2016,Allen2016,
Amet2016,Borzenets2016}. However, in spite of these excellent predispositions to mediate superconductivity, a full control of the supercurrent both in its amplitude and spatial distribution has not been demonstrated up to now. One of the reasons behind this is the difficulty to confine charge carriers in graphene due to the absence of back scattering and Klein tunnelling \cite{katsnelsonbook}. The use of bilayer graphene (BLG) could circumvent these problems since it is possible to engineer an electronic band gap by breaking the lattice inversion symmetry of the AB-stacked bilayer \cite{McCann2006,McCannKoshino2013}. Indeed, by means of local gating, BLG can provide a way to shape the supercurrent distribution and allow a complete monitoring of proximity induced superconductivity. In this work, we have used edge connected BLG-hexagonal boron nitride (hBN) heterostructures as a medium for induced superconductivity, and use a quantum point contact (QPC)-like geometry to study supercurrent confinement. 
	
\begin{figure*}
\includegraphics[width=1\textwidth]{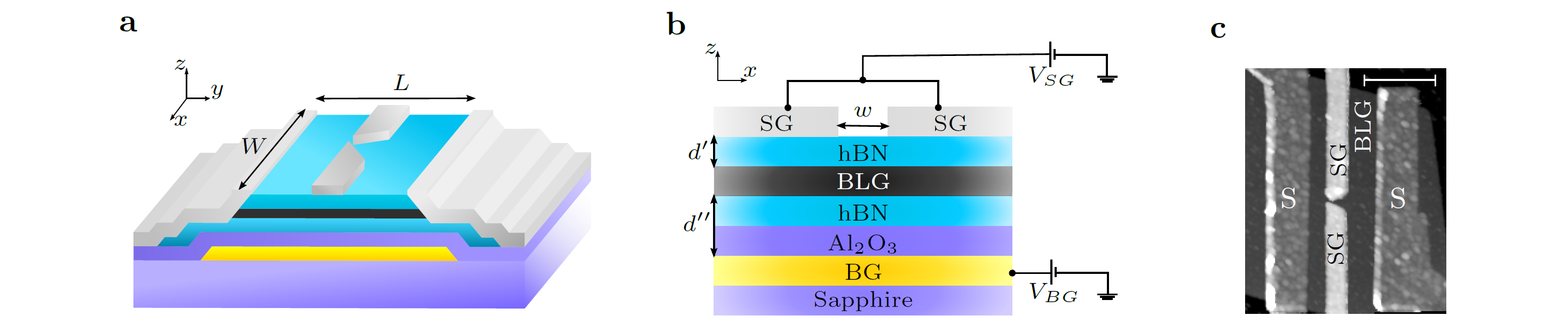}
\caption{\textbf{Device geometry. a,} 3D Schematics of the device and \textbf{b,} cross-sectional view as a cut through the dual-gated region. The device consists of a hBN-BLG-hBN heterostructure on a pre-patterned overall back-gate (BG) and a split-gate (SG). The superconducting leads are edge connected to the mesa. The width \textit{W} = 3.2\,$\mu$m and length \textit{L} = 950\,nm while the distance between the two fingers of the split-gate $w \sim 65\,$nm. \textbf{c,} AFM image of the device. Scale bar is 1\,$\mu$m.}
\label{fig:devicefab}
\end{figure*}

\section{How to read a dual gate map: Inducing a 1D constriction}

The sample geometry used in this study is depicted in Fig.\,1. Following the fabrication method of Wang \textit{et al.} \cite{Wang2013}, we employ BLG encapsulated between hBN multilayers connected from the edge of the mesa with superconducting titanium/aluminium electrodes. The constriction is realised by inducing displacement fields between an overall pre-patterned back-gate and a local top-gate designed in a QPC-like split-gate geometry (see Fig.\,1). Two devices were measured which show similar behaviour, here we present the data based on the shortest sample (details on the sample fabrication are presented in the supplementary information). 
	
\begin{figure*}
\includegraphics[width=1\textwidth]{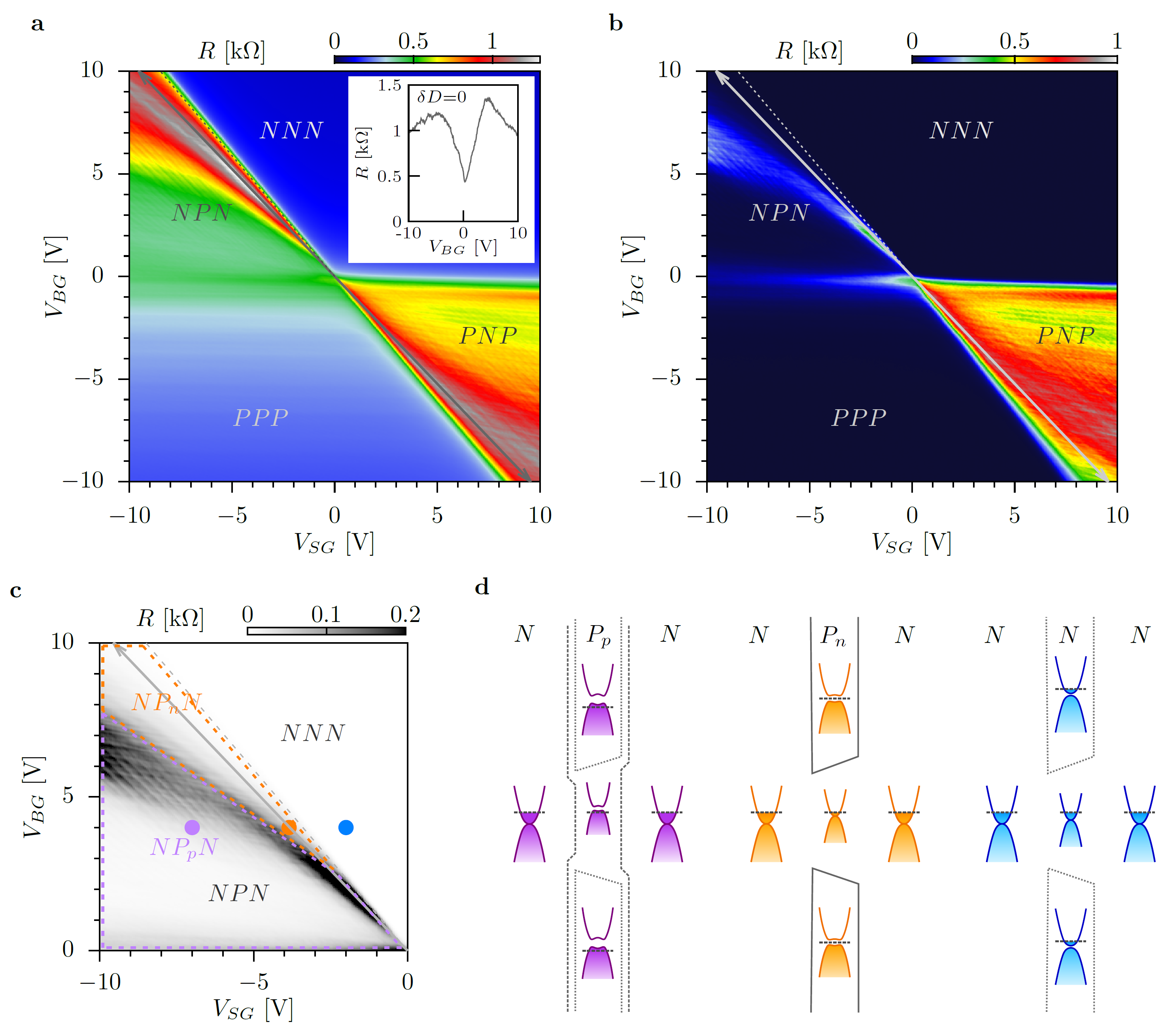}
\caption{\textbf{Formation of the constriction: resistance gate map analysis. a,} Resistance map as a function of back- and split-gate voltage, $V_{BG}$ and $V_{SG}$ respectively, measured at $\sim25\,$mK in the normal state ($B = 20\,$mT). The arrow marks the displacement field line along which the charge carrier density in the dual-gated region is zero. The dashed line indicates the transition when $E_F$ is tuned from the conduction band into the induced band gap, highlighting the crossover to a confined system. The inset displays the normal state resistance measured along the displacement field line. \textbf{b,} Resistance map versus $V_{BG}$ and $V_{SG}$ measured at $\sim25\,$mK in the superconducting state ($B = 0$). \textbf{c,} Zoom-in on the upper left part of the resistance map in the superconducting state (\textbf{b}) where the different regime areas are enlightened, \textit{i.e.} the formed 1D constriction area $NP_{n}N$, the unipolar regime $NNN$ and the non-uniform $NP_{p}N$ junction. \textbf{d,} Schematics of the spatially resolved energy band diagrams of our QPC geometry where top-views of the device refer to the three different regimes of panel \textbf{c}.}
\label{fig:gatemap}
\end{figure*}

\begin{figure*}[htbp]
\includegraphics[width=1\textwidth]{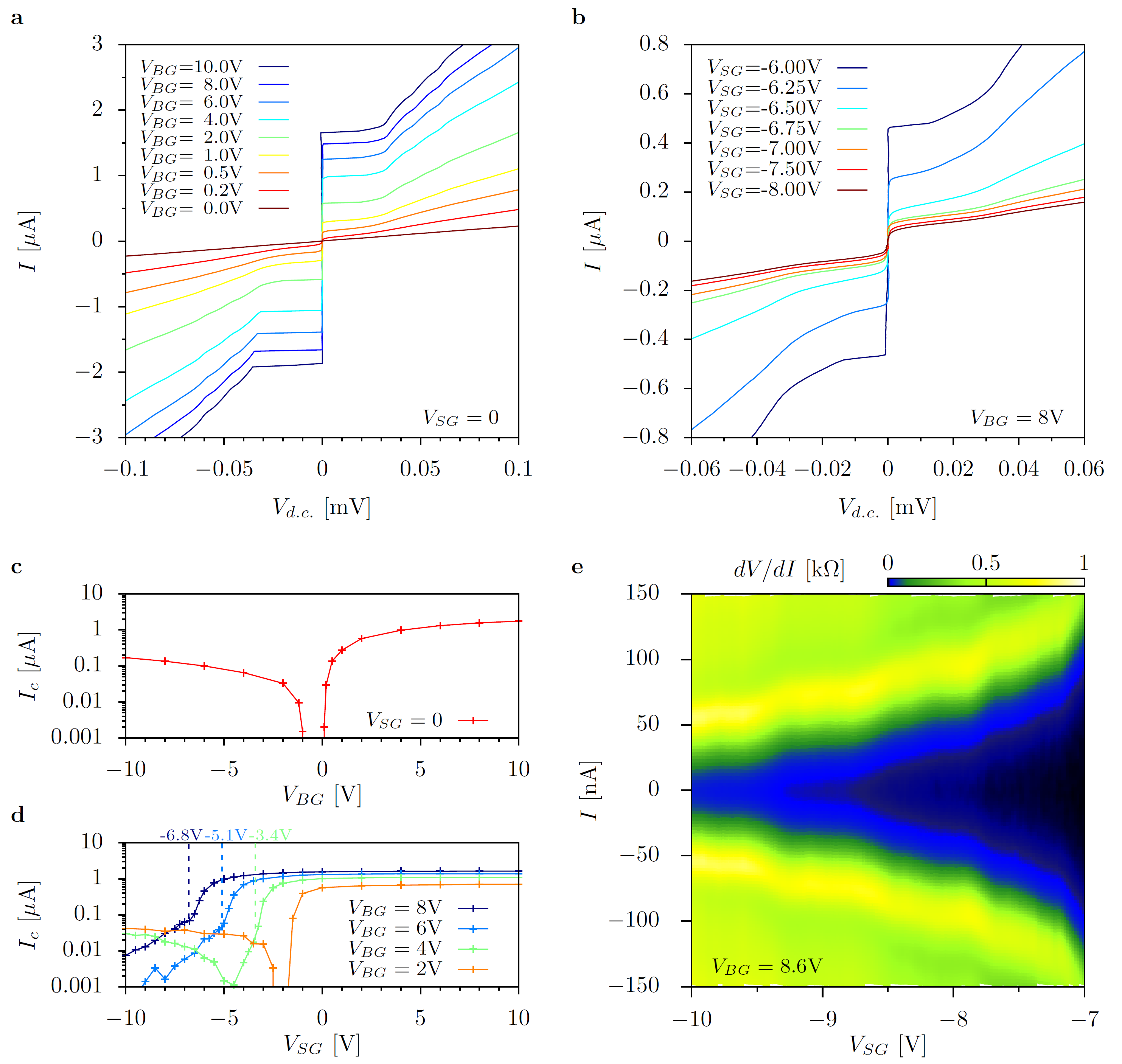}
\caption{\textbf{Gate-controlled current in a superconducting BLG weak link. a,} \textit{I-V} curves for different $V_{BG}$, \textit{i.e.} densities, characterising the 2D system at $V_{SG}=0\,$V. \textbf{b,} \textit{I-V} curves at fixed back-gate voltage $V_{BG}=8\,$V for various displacement fields $D$ in the dual-gated region, \textit{i.e.} for split-gate voltages close to the transition from $NNN$ to $NP_nN$. \textbf{c,} Back-gate voltage dependence $V_{BG}$ of the critical current $I_c$. \textbf{d,} $I_c(V_{SG})$ for constant charge carrier densities (\textit{i.e.} constant $V_{BG}$). \textbf{e,} Resistance map $vs$ $V_{SG}$ and current $I$ zoomed-in on the $NP_nN$ region, revealing a step-wise reduction of the critical current $I_c$.}
\label{fig:supercurrent}
\end{figure*}

The normal state characteristics of our sample show a residual charge carrier density as low as $2.8 \times 10^{10}\,$cm$^{-2}$, well developed Landau fans in magnetotransport experiments as well as multiple Fabry-P\'{e}rot interferences generated by the charge carriers travelling back and forth within the several cavities formed in our system (see supplementary information for the full analysis). Figures\,2a and 2b display resistance maps as a function of split- and back-gate voltage measured in the normal and superconducting state respectively (\textit{i.e.} at 20\,mT and zero magnetic field). In both cases, distinct deviations from the expected quadrants formed in lateral $npn$-junctions corresponding to the differently doped regions \cite{Oostinga2007,Taychatanapat2010,Varlet2014} are clearly visible (unipolar and bipolar regions $NNN$, $PPP$ and $NPN$, $PNP$ respectively). 

In BLG dual-gated devices, the displacement field is used to break the lattice inversion symmetry of the AB-stacked bilayer: the two layers being at different potentials a band gap opens \cite{McCann2006,McCannKoshino2013}, inducing an insulating state with strongly suppressed conductivity. The resistance then raises monotonically with increasing displacement field as the band gap develops \cite{Oostinga2007,Taychatanapat2010,Varlet2014}. Here, we observe a non-monotonic change of the resistance which first increases and then drops after reaching a maximum while following the displacement field line (\textit{i.e.} when the displacement field generated by the back- and split-gates, respectively $D_b$ and $D_t$ are equal, at $\delta D = D_b - D_t = 0$ \cite{Zhang2009}). In addition, the resistance peak does not follow the displacement field line which is indicated by the gray arrow as depicted in Fig.\,2a and 2b, but diverges into the bipolar regions ($NPN$ and $PNP$). This trend is already noticeable in the normal state resistance (Fig.\,\ref{fig:gatemap}a), but becomes strikingly evident in the superconducting state (Fig.\,\ref{fig:gatemap}b). This unexpected behaviour can be understood as the competitive action of back- and split-gates within the constriction. As the displacement field increases, the charge carrier density mostly driven by the back-gate becomes less and less affected by the stray fields developed by the split-gate which cannot compensate the influence of the back-gate on the channel region.  Consequently, the device remains highly conductive in contrast to the pinch-off characteristic of gapped BLG with full-width top-gate. Instead, the maximum resistance deviates from the displacement field line and ``bends''. The bent line of the resistance peak results then from the required overcompensation of the split-gate voltage to diminish the induced charge carriers within the channel region. Instead of being maximum along the displacement field line \cite{Oostinga2007,Taychatanapat2010,Varlet2014}(marked as a diagonal arrowed line on the gate maps), the resistance increases up to a maximum then decreases as plotted in the inset of Fig.\,2a. However, this imbalance between applied split- and back-gate voltages starts to induce charge carriers of opposite sign in the dual-gated cavities, resulting in $pn$-junctions. As a consequence, the bipolar regions become then subdivided into two parts depending on the doping in the constriction (denoted by a sub-label like $NP_nN$, see Fig.\,2c). The QPC-like structure can then be driven in an ``open'' (the 1D channel doping is of the same type as the 2D reservoirs) or ``closed'' (the 1D channel doping is of opposite type as the reservoirs forming a non-uniform potential barrier) regime.

The schematics in Fig.\,\ref{fig:gatemap}d summarize the different scenarios which govern the behaviour of such an electrostatically induced constriction. It is important to note that the overall resistance remains higher on the $p$-side ($PPP$ and $PNP$) due to the slight $n$-doping provided by the leads which create a $pn$-junction at each contact. This becomes particularly clear in the superconducting state where the $PNP$ region remains resistive while a large part of the $NPN$ section displays a zero resistance state. For this reason, we focus on the $NPN$ area and in particular on the $NP_{n}N$ part where we can study the supercurrent flowing through the constriction.    
		
\section{Supercurrent analysis}

Now we describe how to control both supercurrent amplitude and spatial distribution using our split-gate geometry. We have seen in the previous section that our device becomes superconducting in the area where the constriction is formed, namely the $NP_{n}N$ region. One way to verify our hypothesis consists of probing the critical current $I_c$ which corresponds to the maximum supercurrent that a weak link can support before switching to a resistive state (see method section for a description of the critical current extraction procedure and the supplementary information for details). $I_c$ being extremely sensitive to any external perturbations such as magnetic field, potential landscape inhomogeneities or thermal excitation, drastic changes of the confinement should be clearly observed. Indeed, the variation of the normal state resistance is directly reflected in the supercurrent amplitude. For example, small oscillations in the resistance produced by Fabry-P\'{e}rot interferences are directly detected in the supercurrent \cite{Jorgensen2006,Calado2015,BenShalom2016,Borzenets2016} (see supplementary information). Here, we focus our attention on the effect of the 1D constriction on the supercurrent amplitude.  

The amplitude of the supercurrent can be monitored by tuning the charge carrier density with the overall back-gate voltage $V_{BG}$. In Fig.\,3a the current-voltage characteristics are shown in the absence of a constriction, \textit{i.e.} for a uniform 2D weak link at $V_{SG}=0$. The supercurrent evolves from zero at the charge neutrality point up to a measured maximum of $1.86\,\mu$A at high charge carrier density $n=4\cdot 10^{12}\,$cm$^{-2}$ (\textit{i.e.} $V_{BG}=10\,$V). It is important to note that the \textit{I-V} characteristics only display a rather limited hysteretic behaviour visible only at large charge carrier density corresponding to a weakly underdamped junction within the resistively and capacitively shunted junction (RCSJ) model \cite{tinkhambook}. When the Fermi level lies in the valence band ($V_{BG}<0$), the weak link is disturbed by the presence of the $pn$-junctions which strongly suppresses the supercurrent by an order of magnitude (approximately 200\,nA at $V_{BG}=-10\,$V). This is clearly seen in Fig.\,\ref{fig:supercurrent}c where the critical current $I_c$ is plotted as a function of the back-gate voltage $V_{BG}$. 

Fig.\,3b displays a series of \textit{I-V} curves at fixed charge carrier density (here at $V_{BG}=8\,$V) for different split-gate values in the vicinity of the $NP_{n}N$ area. When approaching the formation of the constriction, $I_c$ decreases rapidly until $V_{SG}\sim-6.65\,$V. At this point, the Fermi level underneath the split-gate is positioned in the gap. Therefore, charge carriers can only flow through the 1D constriction. Beyond the formation of the constriction, $I_c$ decreases in a much slower fashion. The extracted critical current $I_c$ is plotted in Fig.\,3d as a function of the split-gate voltage $V_{SG}$ at different densities. At small densities, \textit{i.e.} $V_{BG}=2\,$V (orange curve in Fig.\,3d), the starting point of the $NPN$ region appears early in gate voltage and the supercurrent is switched off. Then, the Fermi level in the constriction which remains mainly driven by the stray fields of the split-gate moves towards the valence band. Due to the close proximity of the split-gates, the stray fields are strong enough to close the channel. A small supercurrent can be detected despite the presence of a weak $pn$-junction as depicted in Fig.\,2d ($NP_pN$ area). In contrast, at higher densities the back-gate starts to electrostatically dominate the constriction region. The creation of the 1D channel is directly reflected in the sudden change of slope of $I_c(V_{SG})$ curves (blue and dark blue curves in Fig.\,3d, the change of slope being marked by dotted lines). The supercurrent through the channel is then only slowly reduced with increasing split-gate voltage owing to the narrowing of the channel by the stray fields. Once the channel is created, the amplitude of the supercurrent drops way below 100\,nA while multiple Andreev reflections completely vanish (see supplementary information). At intermediate density (green curve in Fig.\,3d), the channel is first created (rapid drop in $I_c(V_{SG})$ then change of slope marked by the dotted curve), then closed with the Fermi level positioned in the gap (supercurrent switched off), to finally form a non-uniform $pn$-junction as depicted in Fig.\,2d ($NP_pN$ area).  Importantly, despite the absence of signs of 1D subband formation while shrinking the constriction in the normal state, the critical current decreases in a step-wise fashion (see Fig.\,3e) as predicted for ballistic supercurrents in quantum point contacts \cite{Furasaki1991,Furasaki1992,Takayanagi1995}. 
	
\section{Magneto-interferometry}	

\begin{figure*}
\includegraphics[width=1\textwidth]{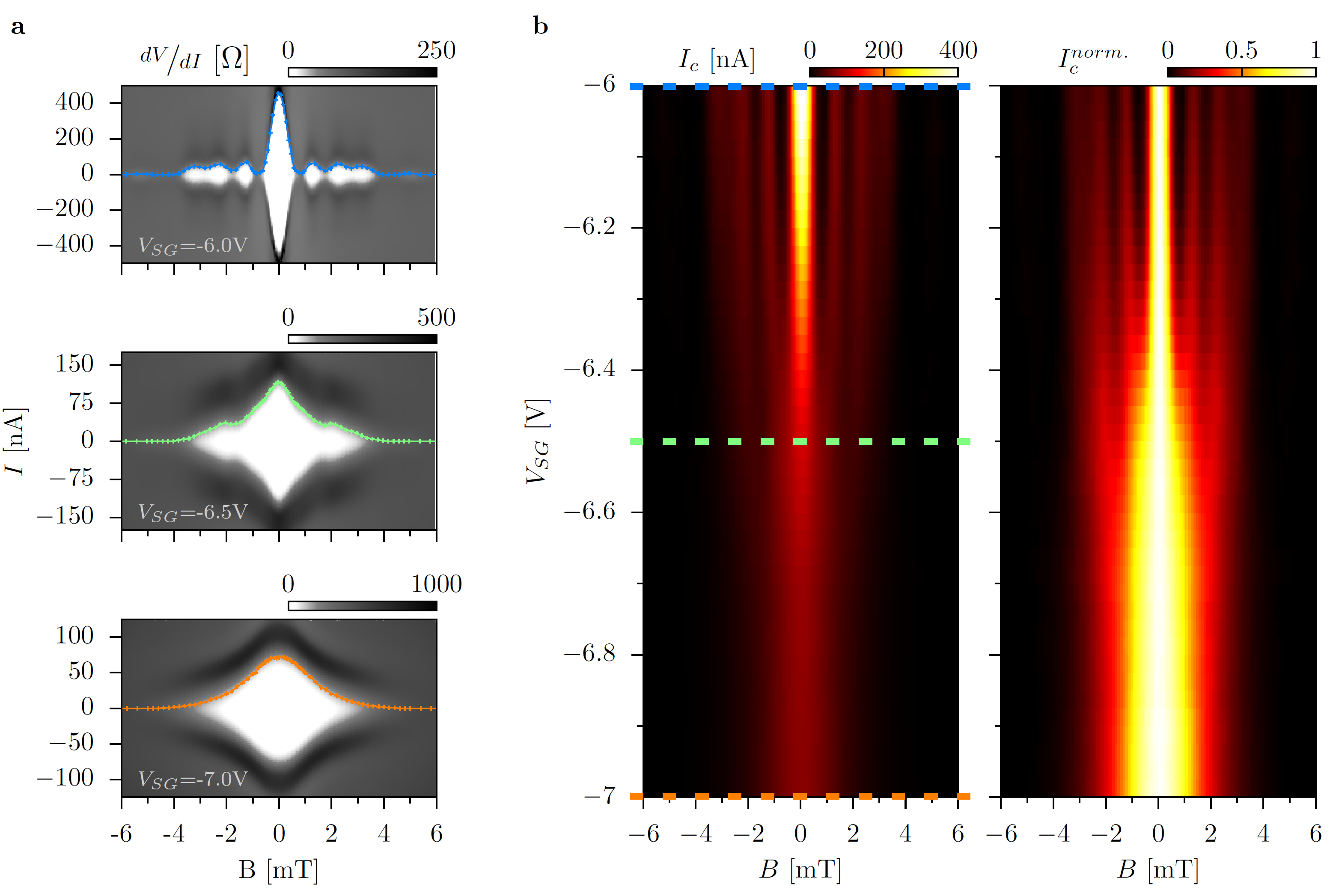}
\caption{\textbf{Magnetic interferometry study of the transition from 2D to 1D confinement of the supercurrent. a,} Gray-scale map of the differential resistance $dV/dI$ versus magnetic field $B$ and current $I$. The coloured dotted lines correspond to the extracted $I_c$. These measurements are taken at three different split-gate voltage values ($V_{SG}$ = 6\,V, 6.5\,V and $7\,$V) at constant charge carrier density ($V_{BG}=8\,$V). Drastic change in the interference pattern is observed highlighting a clear transition from 2D to 1D confined supercurrent. \textbf{b,} Critical current amplitude $I_{c}$ (left panel) and normalised critical current amplitude $I_{c}^{norm.}$ (right panel) mapped as a function of magnetic field $B$ and split-gate voltage $V_{SG}$. The transition from a beating pattern (Fraunhofer-like) to a monotonically decaying pattern is visible confirming the continuous change in the supercurrent confinement from 2D to 1D. The coloured dashed lines correspond to the split-gate values where the $dV/dI(V_{SG},B)$ maps were measured in panels \textbf{a}.}
\label{fig:devicefab}
\end{figure*}

\begin{figure*}
\includegraphics[width=1\textwidth]{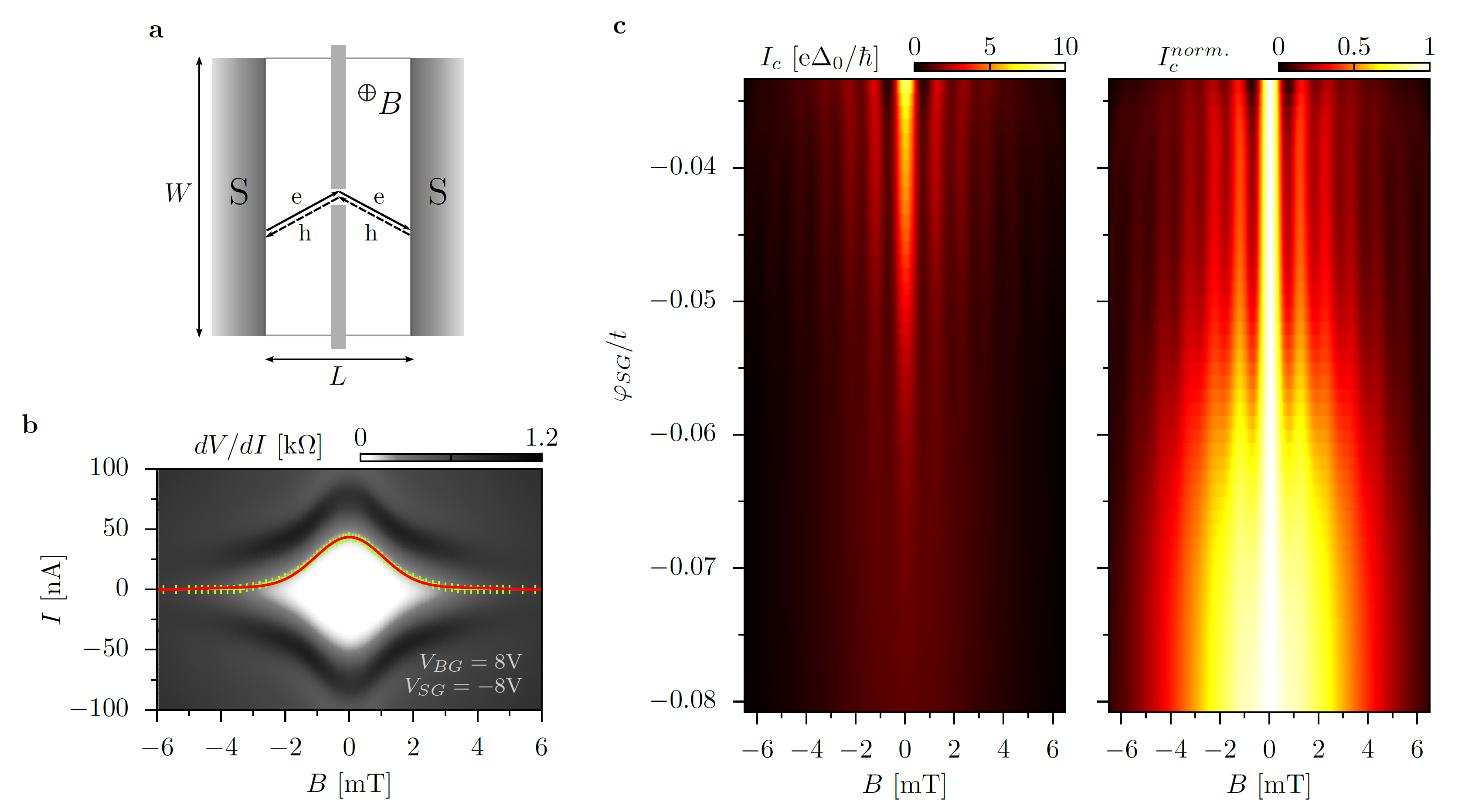}
\caption{\textbf{Modelling supercurrent confinement. a,} Schematic of the superconducting weak link with a quantum point contact like geometry used for our analytical model. \textbf{b,} Differential resistance $dV/dI$ versus  magnetic field $B$ and current $I$ including the extracted critical current $I_c$ (green crosses) fitted with our analytical model (red line) when the 1D constriction is formed (at $V_{BG}=8\,$V and $V_{SG}=-8\,$V). \textbf{c,} Numerical simulations of critical current amplitude $I_{c}$ (left panel) and normalized critical current amplitude $I_{c}^{norm.}$ (right panel) mapped as a function of magnetic field $B$ and split-gate strength $\varphi_{SG}$ showing the transition from 2D to 1D of the magnetic interferences. The $x$-axis is rescaled to magnetic field $B$ using the parameters extracted by fitting the numerical simulation to the experimental data at $V_{SG}$=0 (see supplementary information for details).}
\label{fig:model}
\end{figure*}

The supercurrent density distribution across the sample width can be explored by probing its interference pattern \cite{Rowell1963}  in response to a perpendicular magnetic flux penetrating the junction \cite{DynesFulton1971,Zappe1975,BaronePaternoBook,Barzykin1999,Kikuchi2000,Angers2008,Chiodi2012,Amado2013,Hart2014,Allen2016}. Therefore, by changing the geometry of the system one can observe a large variety of interference patterns directly related to the supercurrent density distribution \cite{BaronePaternoBook}. 
As recently shown \cite{Hart2014, Allen2016}, superconducting interferometry is a powerful tool to probe confinement  where the current density distribution can be extracted by complex Fourier transform following the approach of Dynes and Fulton \cite{DynesFulton1971}. However, this technique of recovering the supercurrent assumes that it is carried strictly in a direction normal to the superconducting electrodes, and therefore does not apply to our device because of its small aspect ratio, especially in the QPC regime.

Here, we show that the magnetic interference pattern indicates clear signatures of the supercurrent confinement. Fig.\,4a exhibits a series of resistance maps versus current and magnetic field at constant density ($V_{BG}=8\,$V). A progressive change of the interference pattern is observed as the split-gate is tuned and the 1D constriction forms. First, a beating pattern appears, resembling Fraunhofer-like interference (upper panel) when the system remains two-dimensional. Then the interference pattern turns to a ``lifting lobes'' shape just before the formation of the constriction (middle panel). Finally a non-beating ``bell-shaped'' pattern is formed while the supercurrent flows only through the confined 1D constriction (lower panel). We note that the transition from a beating to a non-beating pattern occurs on a rather narrow voltage range -7\,V\,$<V_{SG}<$\,-6\,V (at $V_{BG}=8\,$V, additional data at $V_{BG}=4\,$V are shown in the supplementary information). In Fig.\,4b we can observe a map of the critical current $I_c$ (left panel) as well as the critical current normalized with the maximum critical current (at $B=0$) $I_c^{norm.}$ (right panel) as a function of magnetic field $B$ and split-gate voltage $V_{SG}$, allowing a more accurate vision of the transition from 2D (beating pattern) to 1D (``bell-shaped'' pattern). Each horizontal slice of such maps corresponds to the extracted critical current (or normalized critical current) of a single magnetic interference pattern. We note that such non-beating pattern has been observed in rectangular superconducting weak links with low aspect ratio \cite{Angers2008,Chiodi2012,Amado2013}. From the magneto-interferometry experiments, no obvious signs of induced current through topological channels appearing due to AB stacking faults \cite{Ju2015} or edge states \cite{Allen2016,Zhu2017} have been detected.

In order to gain deeper understanding how the magnetic interferences should evolve with the creation of a 1D constriction into a 2D system, we have designed an analytical model where we calculate the Josephson current through the sample in the presence of a magnetic field $B$ (see supplementary information for details), using a quasi-classical approach (as in \cite{Barzykin1999,Sheehy2003,Meier2016}) with an additional input given by the presence of a QPC-like structure in the middle of the device (see the geometry used in Fig.\,5a). We have used our analytic expression to fit the  maximum critical current as a function of magnetic field (see Fig.\,5b). The theoretical critical current (red curve) is matched to the experimental data $I_c(B)$ (green crosses) by scaling the curve by a factor of the extracted maximum critical current $I_c(0)=43.5\,$nA  using a junction area of $\sim 4.81\times 10^{-12}\,$m$^2$ with a total junction length of $\tilde{L}=L+2\lambda_L=1.50\,\mu$m where $\lambda_L$ is the London penetration depth ($\lambda_L \sim 275\,$nm). Our model follows clearly the experimental data $I_c(B)$ which, once again, proves that the supercurrent has been strongly confined in our quantum point contact edge connected BLG. We finally show tight-binding simulations using Kwant package \cite{Groth14} of $I_c$ as a function of magnetic field $B$ and split-gate strength $\varphi_{SG}$ in Fig.\,5c (see supplementary information for details) which are in good qualitative agreement with our experimental data of Fig.\,4c.   
 
	
\section{Conclusion and perspectives}	

In this work, we have demonstrated a full monitoring, both spatially and in amplitude, of the supercurrent in a clean and edge connected hBN-BLG-hBN heterostructure. In a split-gate geometry we have explored the consequences of the 1D confinement on the supercurrent and on its magnetic interferences. Thanks to in turn, the possibility to locally engineer an electronic band gap in BLG, the injection of a large and fully tunable critical current, and the ultra-low disorder of fully encapsulated hBN-BLG-hBN heterostructures, we have designed a unique platform allowing the creation of new types of superconducting circuits based on fully tunable weak links which can be controlled by the combination of top- and back-gates.


\vspace{5mm}

\textbf{Acknowledgements}
\\
The authors thank A. Mirlin, M. Titov and W. Wernsdorfer for fruitful discussions. This work was partly supported by Helmholtz society through program STN and the DFG via the projects DA 1280/3-1 and GO 1405/3-1.
A.A. and M.I. acknowledge support of the European Research Council, and the Netherlands Organisation for Scientific Research (NWO/OCW), as part of the Frontiers of Nanoscience program.

\vspace{5mm}

\textbf{Author contributions}
\\
R.Kra. performed the experiments with the support of J.M., R.Du., P.B.S., F.W. and R.Da. R.Kra. fabricated the devices with the support of J.M. U.N.K. and I.G. designed the analytical model. M.I. and A.A. performed the numerical calculations. All authors discussed about the results. R.Da. and R.Kra. performed the data analysis and wrote the paper. R.Da. designed and planned the experiments. 

\vspace{5mm}

\textbf{Additional information}

Correspondence and requests for materials should be addressed to R.Da. (e-mail: romain.danneau@kit.edu)

\vspace{5mm}

\textbf{Competing financial interests}

The authors declare no competing financial interests.

\vspace{5mm}

\subsection{Method subsection.}

\textit{Experimental}: The low-temperature electrical measurements were performed in a Bluefors LD250 $^{3}$He/$^{4}$He dilution fridge. The base temperature of the measurement was about 25 mK. All dc-lines were strongly filtered using 3-stage RC-filters with a cut-off frequency of 1 kHz, as well as PCB-powder filters with a cut-off frequency of about 1 GHz. The differential resistance/conductance data was measured using standard low-frequency ($\sim$13 Hz) and various low excitation (between 1 and 10 $\mu$V), the gating and the out-of-equilibrium measurements were performed using ultra-low noise dc-power supply from Itest. The normal state was obtained by applying a perpendicular magnetic field of 20 mT. The experiments were performed within several thermal cycles (room temperature $\rightleftharpoons$ milli-Kelvin temperature). Data have been reproduced and implemented in each cooldown.

\textit{Data treatment and $I_c$ extraction:}
The critical current $I_c$ is extracted using a voltage threshold method, where the threshold is set to 1~$\mu$V. The two adjacent data points of recorded IVs right before and after the threshold are evaluated and $I_c$ is determined by linear extrapolation in the current of these two points depending on the difference of the voltage drop with respect to the threshold. The extracted critical current is corrected by subtracting the artificial offset that is produced by this method. 


\end{document}


\title{SUPPLEMENTARY INFORMATION: Tailoring supercurrent confinement in graphene bilayer weak links}

\author{Rainer Kraft}
\affiliation{Institute of Nanotechnology, Karlsruhe Institute of Technology, D-76021 Karlsruhe, Germany}

\author{Jens Mohrmann}
\affiliation{Institute of Nanotechnology, Karlsruhe Institute of Technology, D-76021 Karlsruhe, Germany}

\author{Renjun Du}
\affiliation{Institute of Nanotechnology, Karlsruhe Institute of Technology, D-76021 Karlsruhe, Germany}

\author{Pranauv Balaji Selvasundaram}
\affiliation{Institute of Nanotechnology, Karlsruhe Institute of Technology, D-76021 Karlsruhe, Germany}
\affiliation{Department of Materials and Earth Sciences, Technical University Darmstadt, Darmstadt, Germany}

\author{Muhammad Irfan}
\affiliation{Kavli  Institute  of  Nanoscience,  Delft  University  of  Technology,
P.O.  Box  4056,  2600  GA  Delft,  The  Netherlands}

\author{Umut Nefta Kanilmaz}
\affiliation{Institute of Nanotechnology, Karlsruhe Institute of Technology, D-76021 Karlsruhe, Germany}
\affiliation{Institute for Condensed Matter Theory, Karlsruhe Institute of Technology, D-76128 Karlsruhe, Germany}

\author{Fan Wu}
\affiliation{Institute of Nanotechnology, Karlsruhe Institute of Technology, D-76021 Karlsruhe, Germany}
\affiliation{College of Optoelectronic Science and Engineering, National University
of Defense Technology, Changsha 410073, China}

\author{Detlef Beckmann}
\affiliation{Institute of Nanotechnology, Karlsruhe Institute of Technology, D-76021 Karlsruhe, Germany}

\author{Hilbert von L\"{o}hneysen}
\affiliation{Institute of Nanotechnology, Karlsruhe Institute of Technology, D-76021 Karlsruhe, Germany}
\affiliation{Institute of Physics, Karlsruhe Institute of Technology, D-76049 Karlsruhe, Germany}
\affiliation{Institute for Solid State Physics, Karlsruhe Institute of Technology, D-76021 Karlsruhe, Germany}


\author{Ralph Krupke}
\affiliation{Institute of Nanotechnology, Karlsruhe Institute of Technology, D-76021 Karlsruhe, Germany}
\affiliation{Department of Materials and Earth Sciences, Technical University Darmstadt, Darmstadt, Germany}

\author{Anton Akhmerov}
\affiliation{Kavli  Institute  of  Nanoscience,  Delft  University  of  Technology,
P.O.  Box  4056,  2600  GA  Delft,  The  Netherlands}

\author{Igor Gornyi}
\affiliation{Institute of Nanotechnology, Karlsruhe Institute of Technology, D-76021 Karlsruhe, Germany}
\affiliation{Institute for Condensed Matter Theory, Karlsruhe Institute of Technology, D-76128 Karlsruhe, Germany}
\affiliation{A.F. Ioffe Physico-Technical Institute, 194021 St. Petersburg, Russia}

\author{Romain Danneau}
\affiliation{Institute of Nanotechnology, Karlsruhe Institute of Technology, D-76021 Karlsruhe, Germany}

\maketitle


\section{Device fabrication}

The edge connected van der Waals heterostructures based on hBN-bilayer graphene-hBN are prepared following the method developed by Wang \textit{et al.} \cite{Wang2013}. Bilayer graphene (BLG) flakes and hexagonal boron nitride multilayers (bottom and top hBN multilayer are $\sim 35\,$nm and $\sim 38\,$nm thick respectively) are obtained by mechanical exfoliation from natural bulk graphite (NGS Naturgraphit GmbH) and commercial hBN powder (Momentive, grade PT110) respectively and transferred on $p$-doped Si substrates with $300\,$nm thick thermally grown SiO$_{2}$ layer and selected by optical contrast \cite{Blake2007}. Raman spectroscopy  was used to unambiguously identify BLG \cite{Ferrari2006} (Renishaw inVia Raman spectrometer, using a laser of wave length $\lambda= 532\,$nm). The graphene was then encapsulated between a top and a bottom hBN flake by piling up the layers sequentially by using a polymer-free assembly technique \cite{Wang2013} and a home-made transfer set-up. Then, the whole stack was transferred onto a sapphire substrate with a Cr/Au ($5\,$nm/50$\,$nm) pre-patterned back-gate which is covered by a dielectric of $20\,$nm Al$_{2}$O$_{3}$ deposited by atomic layer deposition \cite{Benz2013} (182 cycles at 200$\,^{\circ}$C). Edge contacts are designed on the mesa and defined by electron beam lithography using a single resist layer of PMMA covered by conductive polymers (Espacer 300Z from Showa Denko K.K., see \cite{Pallecchi2011}) for both etching and subsequent metallisation. The conductive polymer insures good evacuation of charges allowing e-beam lithography on a fully insulating substrate such as sapphire. Unlike in \cite{Wang2013}, only PMMA was used as a mask ($\sim 250\,$nm). The hBN-BLG-hBN sandwich was then etched in an Oxford Instruments Plasmalab 80 reactor with a mixture of CHF$_3$ and O$_2$ forming a 60\,W plasma (40\,sccm CHF$_3$ with 4\,sccm O$_2$ at a pressure of 60\,mTorr; etching rate: 23 and 48\,nm/min for PMMA and hBN respectively). A double layer of titanium/aluminium (5\,nm/80\,nm) electrodes were then deposited by molecular beam epitaxy (at  pressure and sample temperature $\sim 10^{-10}$\,mTorr and $\sim -130\,^{\circ}$C respectively) using the same already patterned PMMA resist, followed by lift-off in acetone. In a subsequent step, split-gates are fabricated in the fashion, \textit{i.e.} e-beam lithography followed by metallisation (a Ti/Al double layer of $5\,$nm and $80\,$nm thickness respectively). It is important to note that the width of the split-gates ($\sim 300$\,nm) does not exceed two times the London penetration depth ($\lambda_L > 200\,$nm in aluminium thin films according to \cite{Steinberg2008}) so that the applied magnetic field is not disturbed by the split-gate electrodes becoming superconducting at very low temperature. 
Finally, the devices are shaped into the desired geometry by a third lithography step and subsequent etching using the parameters of the previous etch process. The distance between the two fingers of the split-gate is $w \sim 65\,$nm.

\begin{figure*}
\includegraphics[width=1\textwidth]{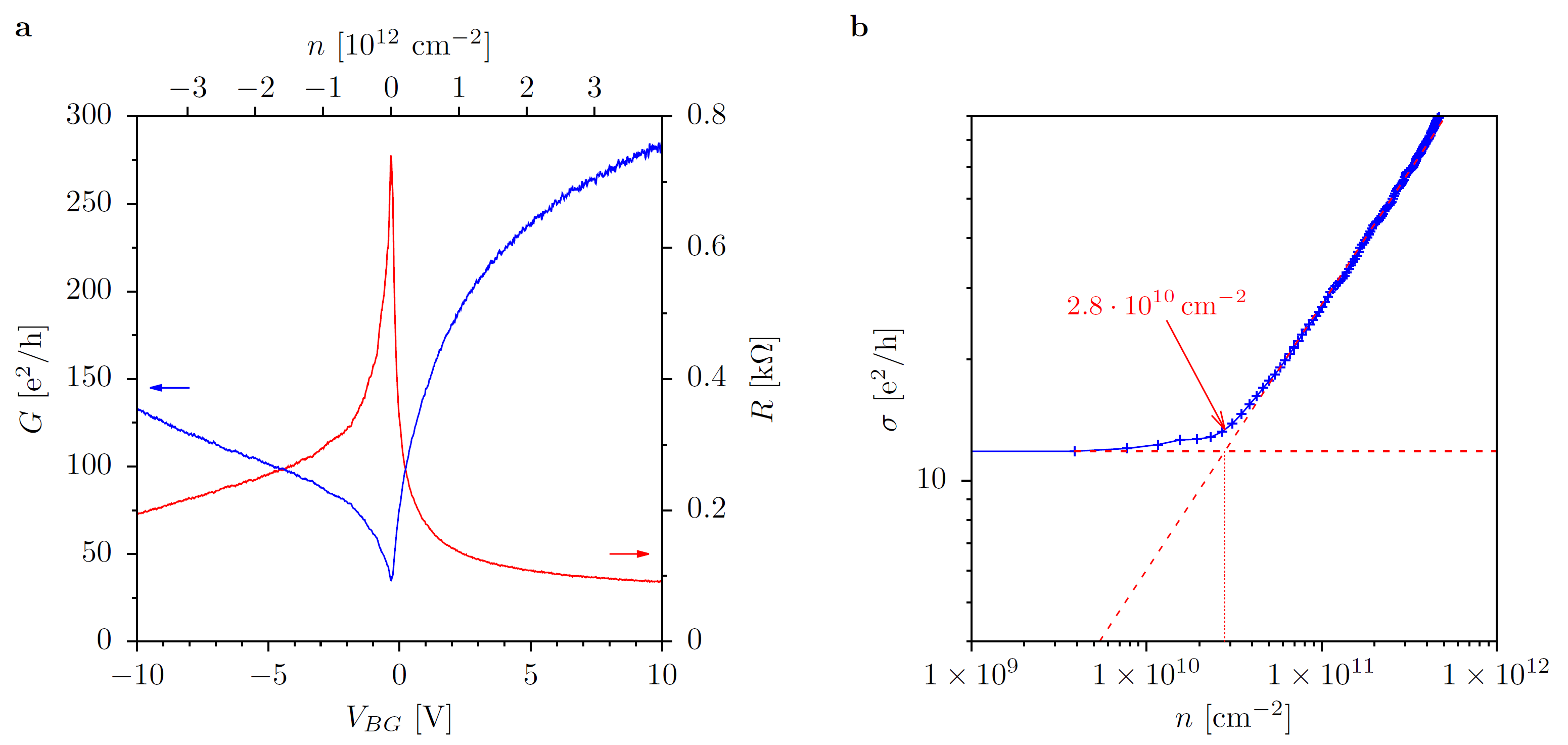}
\caption{\textbf{Normal state characterization. a,} Resistance $R$ (red curve) and conductance $G$ (blue curve) as a function of the back-gate voltage $V_{BG}$ (lower abscissa) or converted charge carrier density $n$ (upper abscissa) of the 2D system respectively. The curve was measured in a separate cooldown at a split-gate voltage $V_{SG}=V_{SG}^{(0)}=0.2\,$V. \textbf{b,} log-log-plot of the conductivity $\sigma$ as a function of $n$. A contact resistance of $90\,\Omega$ was subtracted. The crossing of the saturation conductivity with the extrapolated linear fit to the data at high charge carrier density yields a residual charge inhomogeneity of the order of $10^{10}\,$cm$^{-2}$.  It is important to note that clear Fabry-P\'{e}rot resonances (see Fig.\,\ref{fig:FP}) are visible in all cavities formed by the split-gate or by the unintentional doping from the leads, indicating ballistic transport across the devices.}
\label{fig:resmob}
\end{figure*}

\begin{figure*}
\includegraphics[width=1\textwidth]{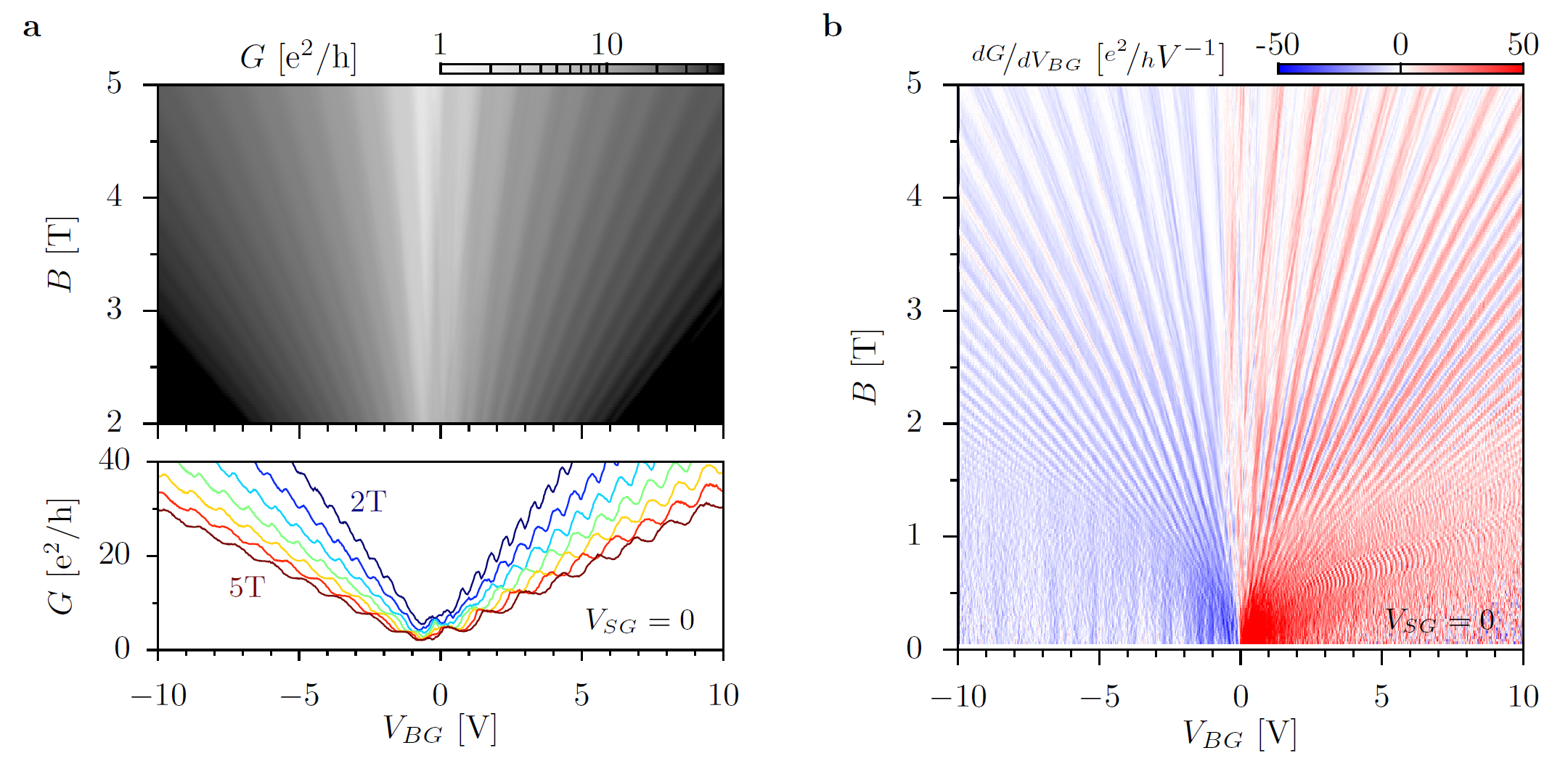}
\caption{\textbf{Two-terminal magnetotransport of the 2D bilayer graphene.} Landau fan diagram of, \textbf{a}, the two-terminal conductance $G(B,V_{BG})$ and \textbf{b}, the two-terminal differentiated conductance $dG/dV_{BG}(B, V_{BG})$ as a function of the magnetic field $B$ and the back-gate voltage $V_{BG}$, measured at $T \sim 25\,$mK.}
\label{fig:landau}
\end{figure*}

\section{Normal state characterization}


Fig.\,\ref{fig:resmob}a shows both normal state resistance $R$ and conductance $G$ as a function of back-gate voltage $V_{BG}$ and charge carrier density $n$, while Fig.\,\ref{fig:resmob}b displays the electron conductivity (minus the estimated contact resistance calculated below) \textit{vs} charge carrier density.

We have extracted the contact resistance from the normal state data at high charge carrier density (\textit{i.e.} $n\sim 4\cdot10^{12}\,$cm$^{-2}$) to insure that the measured normal state resistance $dV/dI=90\,\Omega$ is mostly associated with the contact interfaces. We estimate the number of channels $N=4W/(\lambda_F/2)=4W\sqrt{(n/\pi)}=144$ where the factor $4$ accounts for the spin and valley degeneracy, $W$ is the width of the graphene sheet and $\lambda_F$ the Fermi wavelength. The contact resistivity is then $\rho_c=115\,\Omega\mu$m, comparable to the values given by Wang \textit{et al.} \cite{Wang2013}. 

We estimate the residual density of $n\approx2.8\cdot10^{10}\,$cm$^{-2}$ as to \cite{Du2008} on the electron side (see Fig.\,\ref{fig:resmob}a). Fig.\,\ref{fig:landau}a displays a gray-scale map of the two-terminal conductance $G(B,V_{BG})$ (upper panel) and $G(V_{BG})$ curves for various $B$ (lower panel). The two-terminal differentiated conductance $dG/dV_{BG}(B, V_{BG})$ map as a function of the magnetic field $B$ and the back-gate voltage $V_{BG}$ is shown in Fig.\,\ref{fig:landau}b. The two-terminal quantum Hall conductance curves display  distorted conductance plateaus as expected for high aspect ratio $W/L$ sample geometry \cite{Abanin2008,Williams2009,Du2009,Bolotin2009,Skachko2010}. One can observe a well developed Landau fan  even at relatively low magnetic field, highlighting the high quality of our device.

\begin{figure*}
\includegraphics[width=1\textwidth]{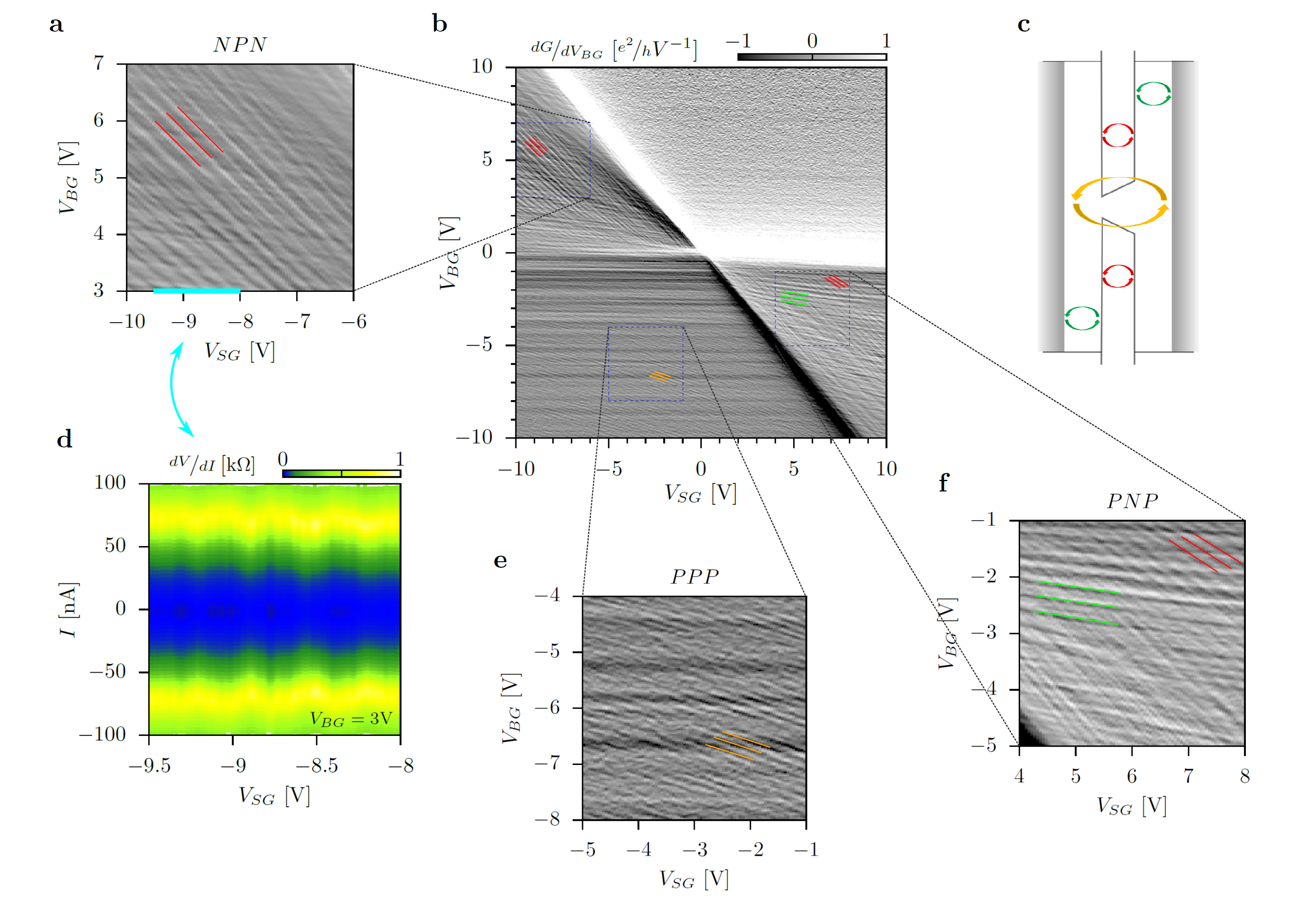}
\caption{\textbf{Fabry-P\'{e}rot resonances. b,} Differentiated conductance $dG/dV_{BG}$ as a function of back- and split-gate voltage ($V_{BG}, V_{SG}$) measured at $20\,$mT (normal state). Fabry-P\'{e}rot resonances can be observed in the parts of the map where cavities are formed by $pn$-junctions. \textbf{a, e, f,} Zoom-in on the $NPN, PNP$ and $PPP$ region of the gate map. The visible resonances are highlighted by exemplary lines and schematically represented on panel \textbf{c,} where the device can be divided into the corresponding cavities. \textbf{d}, differential resistance $dV/dI$ $vs$ current bias $I$ and split-gate voltage $V_{SG}$ measured along the cyan line of panel \textbf{a}. Oscillations of the supercurrent due to the Fabry-P\'{e}rot interferences are visible.}
\label{fig:FP}
\end{figure*}

\section{Fabry-P\'{e}rot interference analysis}

		The gate dependence of the conductance reveals multiple oscillation patterns that can be attributed to Fabry-P\'{e}rot interferences of different cavities \cite{Shytov2008,Young2009,Rickhaus2013,Varlet2014}. Fig.\,\ref{fig:FP} shows $dG/dV_{BG}$ as a function of back- and split-gate voltage ($V_{BG},\,V_{SG}$), where the data of the conductance $G$ is numerically differentiated with respect to $V_{BG}$. Taking into account the size of the cavity and the gate dependence of the interferences, \textit{i.e.} slope and appearance in the gate map, the observed patterns can be unambiguously assigned to the corresponding cavities (see Fig.\,\ref{fig:FP}a, c, e, f). The effective cavity length $L$ can be determined by the relation $\Delta n=2\sqrt{\pi n} /L$ which follows from the resonance condition $\Delta k\cdot L=\pi$.

In the unipolar regime $PPP$, $pn$-junctions arise at the interface of the graphene sheet with the two metallic electrodes. Thus, a cavity is formed between the two outer contacts. At a density $n\approx2\cdot 10^{12}\,$cm$^{-2}$ ($V_{SG}=0$) the spacing between interference peaks is $\Delta n\approx 5\cdot 10^{10}\,$cm$^{-2}$. By using $\Delta n=2\sqrt{\pi n} /L$, we find an effective cavity length of $L\approx 1\,\mu$m which is consistent with the geometrical size of the device. Therefore, we can conclude that such Fabry-P\'{e}rot interferences indicate ballistic transport on a length scale of at least twice the device length, \textit{i.e.} $\approx 2\,\mu$m. 

\begin{figure*}
\includegraphics[width=1\textwidth]{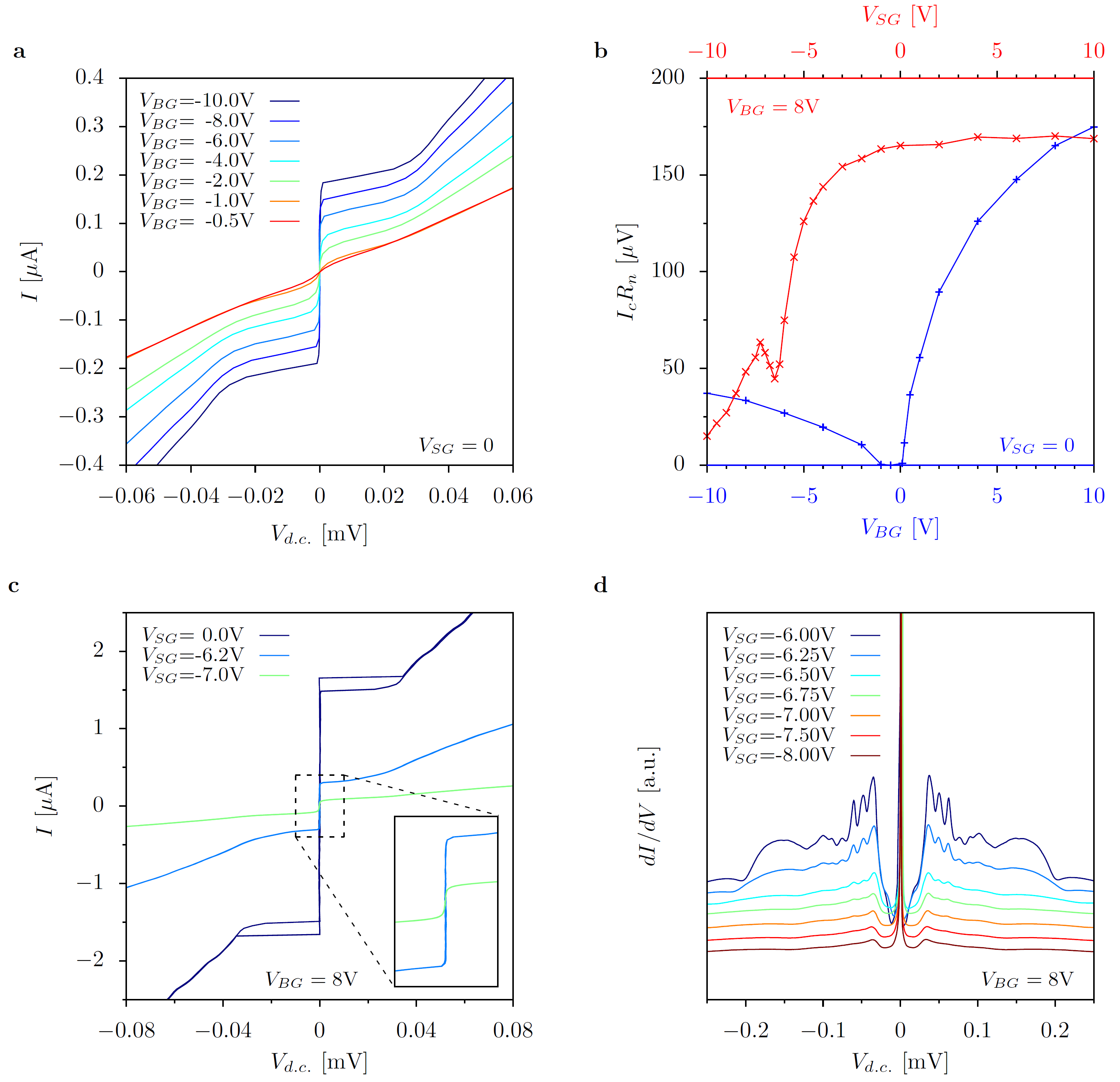}
\caption{\textbf{\textit{I-V} curves, $I_cR_n$ product and multiple Andreev reflections. a,} Series of \textit{I-V} curves in the p-doped region at $V_{SG}=0$. \textbf{b,} $I_cR_n$ versus $V_{BG}$ at $V_{SG}=0$ (in blue) and $I_cR_n$ versus $V_{SG}$ at $V_{BG}=8\,$V (in red). \textbf{c,} The \textit{I-V} curves show the behaviour for the three different cases: no split-gate ($V_{SG}=0$),  right before ($V_{SG}=-6.2\,$V) and once the constriction is fully developed ($V_{SG}=-7\,$V), at $V_{BG}=8\,$V. A zoom-in of the two curves taken at $V_{SG}$=-6.2\,V and -7\,V is displayed in the inset, highlighting the absence of hysteresis before and after the formation of the constriction. \textbf{d,} Series of differential conductance $dI/dV$ $vs$ voltage $V_{d.c.}$ for various split-gate voltages $V_{SG}$ at $V_{BG}=8\,$V. The curves are shifted for clarity. MAR clearly vanishes with the creation of the 1D constriction.}
\label{extrasuper}
\end{figure*}

\section{Additional superconductivity data}

	\subsection{Supercurrent and multiple Andreev reflection}

	Here, we provide additional information on the superconducting properties of our sample. As we have seen, the magnitude of the supercurrent is back-gate tunable, \textit{i.e.} depends on the charge carrier density. While at the charge neutrality point superconductivity is fully suppressed, large supercurrent densities of $\sim 580\,$nA$\,\mu$m$^{-1}$ at $n\approx 4\cdot 10^{12}\,$cm$^{-2}$ are measured. Furthermore, the supercurrent amplitude is dramatically reduced by the two $pn$-junctions formed at bilayer graphene-metal contact interfaces when the back-gate tunes the Fermi level in the valence band ($V_{BG}<0$) (\textit{i.e.} the \textit{PPP} and \textit{PNP} regions). The current-voltage characteristics extracted from the $PPP$ region show a strong attenuation (by approximately one order of magnitude) compared to the $NNN$ part where no $pn$-junction is formed (see Fig.\,\ref{extrasuper}a). The product of the critical current and the normal state resistance $I_cR_n$ as a function of the back-gate voltage $V_{BG}$ at $V_{SG}=0$ (in blue) and $I_cR_n$ as a function of the split-gate voltage $V_{SG}$ at $V_{BG}=8\,$V (in red) are displayed in Fig.\,\ref{extrasuper}b. In the absence of voltage applied on the split-gate, a clear asymmetry between hole and electron conduction is visible in the bipolar supercurrent reflecting the presence of the two $pn$-junctions formed at the contacts. When the split-gate voltage increases, the $I_cR_n$ product remains stable until the constriction starts to form. Then, the $I_cR_n$ product decreases rapidly. We note that contrary to $I_c(V_{SG})$ which decays monotonically, the $I_cR_n$ product suddenly increases slightly between 6\,V and 7\,V corresponding to the voltage range of the 2D to 1D transition, before decreasing again as the constriction size shrinks.   	

{\renewcommand{\arraystretch}{2}%
\begin{table}[h]
\begin{tabular}{|c|c|c|c|c|c|c|c|c|}
  \cline{2-8}
  \multicolumn{1}{c|}{} & $V_{SG} \left[V \right]$ & $V_{BG} \left[ \mathrm{V} \right] $ & $n \left[ \mathrm{cm}^{-2} \right] $ & $I_{c} \left[ \mu \mathrm{A} \right] $ & $I_{r} \left[ \mu \mathrm{A} \right] $ & $I_{r}/I_{c}$ & $I_{c}R_{n}\left[ \mu \mathrm{V} \right] $\\
\hline
  \multirow{5}{*}{\textit{NNN}} & 0 & 10 & $\sim 4 \times 10^{12}$ & 1.86 & 1.66 & 0.9 & 174.8 \\    \cline{2-8}
    & 0 & 8 & $\sim 3.2 \times 10^{12}$ & 1.66 & 1.48 & 0.9 & 165.1 \\
    \cline{2-8}
    & 0 & 6 & $\sim 2.4 \times 10^{12}$ & 1.38 & 1.25 & 0.9 & 147.7 \\
    \cline{2-8}
    & 0 & 4 & $\sim 1.6 \times 10^{12}$ & 1.05 & 0.95 & 0.9 & 126 \\
    \cline{2-8}
    & 0 & 2 & $\sim 0.8 \times 10^{12}$ & 0.585 & 0.585 & 1 & 89.5 \\ \hline
  \multirow{2}{*}{\textit{PPP} } & 0 & -10 & $\sim -4 \times 10^{12}$ & 0.185 & 0.185 & 1 & 37.2 \\
  \cline{2-8}
    & 0 & -8 & $\sim -3.2 \times 10^{12}$ & 0.152 & 0.152 & 1 & 33.4 \\ \hline
  \textit{NNN} & -6.2 & 8 & $\sim 3.2 \times 10^{12}$ & 0.295 & 0.295 & 1 & 56.3 \\ \hline
 \textit{NP$_n$N} & -7 & 8 & $\sim 3.2 \times 10^{12}$ & 0.071 & 0.071 & 1 & 58.1 \\ \hline
\end{tabular}
\caption[]{Critical current $I_c$, the retrapping current $I_r$, $I_r/I_c$ and $I_{c}R_{n}$ product under several gate conditions corresponding to different regions of the gate map (see Fig.\,2 of the main text).}
\label{table:name}
\end{table}}

\begin{figure*}
\includegraphics[width=1\textwidth]{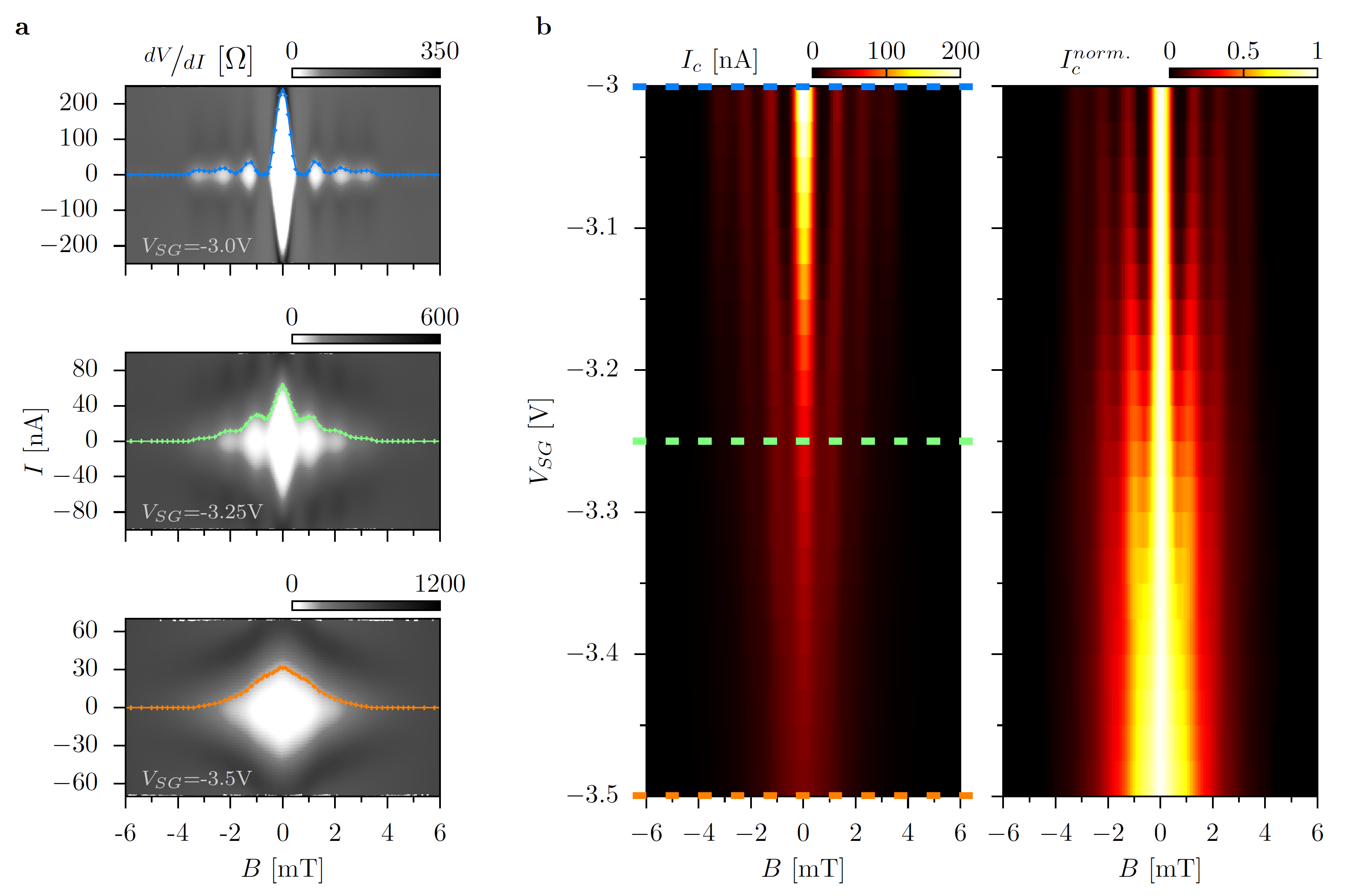}
\caption{\textbf{Magnetic interferometry study of the transition from 2D to 1D confinement of the supercurrent at $V_{BG}=4\,$V. a,} Gray-scale maps of the differential resistance $dV/dI$ versus bias current $I$ and magnetic field $B$ measured at $V_{SG}=-3\,$V, $V_{SG}=-3.25\,$V and $V_{SG}=-3.5\,$V. The coloured doted lines correspond to the extracted $I_c$. \textbf{b,} Critical current $I_{c}$ (left panel) and normalized critical current $I_c^{norm.}$ (right panel) mapped as a function of magnetic field $B$ and split-gate voltage $V_{SG}$. The coloured dashed lines correspond to side-gate values where the $dV/dI(B,I)$ maps were measured in panels \textbf{a}.}
\label{Maginter}
\end{figure*}

	The impact of the constriction on the hysteretic behaviour of the Josephson effect is shown in Fig.\,\ref{extrasuper}c. As we can see on the \textit{I-V} curves (up and back bias sweeps), the hysteresis occurs at sufficiently high charge carrier density and disappears once the constriction develops. Table\,\ref{table:name} recapitulates the extracted $I_c$, the retrapping current $I_r$ and the ratio $I_r/I_c$ in the various regions of the gate map (see Fig.\,2 of the main text). Within the RCSJ model \cite{tinkhambook}, the Josephson junction is tuned from underdamped to overdamped. We note that for small n-doping, \textit{i.e.} $n < 1.5\cdot10^{12}\,$cm$^{-2}$, as well as for $p$-doping no hysteresis is detected.
	
	Fig.\,\ref{extrasuper}d exhibits the effect of the constriction on multiple Andreev reflections (MAR). MAR appear as peaks in the differential conductance $dI/dV$ versus bias voltage $V_{d.c.}$ and are positioned at $2\Delta/en$ (with $n=1,2,3,4,...$ and  $\Delta$ being the superconducting gap, here estimated at $\sim\,100\,\mu$eV). When the constriction is formed, these subgap features disappear while a finite supercurrent remains detectable. The constriction limits dramatically the possibilities for the reflected quasi-particles to return to the opposite lead, and therefore vanishes the MAR. It is important to note that the disappearance of MAR coincides with the reduction of the supercurrent amplitude and the change in the magneto-interferometric pattern (see Fig.\,3d and 4 in the main text). 

	
	
	
\subsection{Magnetic interference patterns at $V_{BG}=4\,$V}

Here we show an additional series of data where the change of the magnetic interferences clearly shows a transition from a beating to a non-beating pattern corresponding to the creation of the 1D constriction. Fig.\,\ref{Maginter}a displays a series of resistance maps versus current and magnetic field at constant density ($V_{BG}=4\,$V) in a similar fashion as the data taken at higher density presented in the main text. Here, the transition from a beating to a non-beating pattern occurs in a voltage range -3\,V $<V_{SG}<$ -3.5\,V. In Fig.\,\ref{Maginter}b, we see the two coloured maps of the critical current $I_c$ (left panel) and the critical current normalized with the maximum critical current (at $B = 0$) $I_c^{norm.}$ (right panel) as a function of magnetic field $B$ and split-gate voltage $V_{SG}$.

\subsection{Analytical model: Long junction }

We calculate the Josephson current $J(\chi)$  through the sample as a function of the superconducting phase difference $\chi=\chi_2-\chi_1$ 
in the presence of magnetic field $B$, using the quasiclassical approach developed in Refs.~\cite{Barzykin1999,Sheehy2003} and \cite{Meier2016}.
The essence of this approach lies in expressing the superconducting current density in terms of quasiclassical
trajectories  connecting the superconducting leads. These paths can be viewed as electron-hole ``tubes'' of width $\sim \lambda_F$, resulting from the Andreev reflection at the NS interfaces and corresponding to Andreev bound states. Each path is associated with the partial contribution to the Josephson current that
depends on the positions of end points ($y_1,y_2$ in Fig.~\ref{Fig:1}) and on magnetic field that enters through the Aharonov-Bohm phase.

The magnetic interference pattern is obtained after the summation over all paths and maximizing the Josephson current with respect to $\chi$:

\begin{figure}
  \centering
  \includegraphics[width=.35\textwidth]{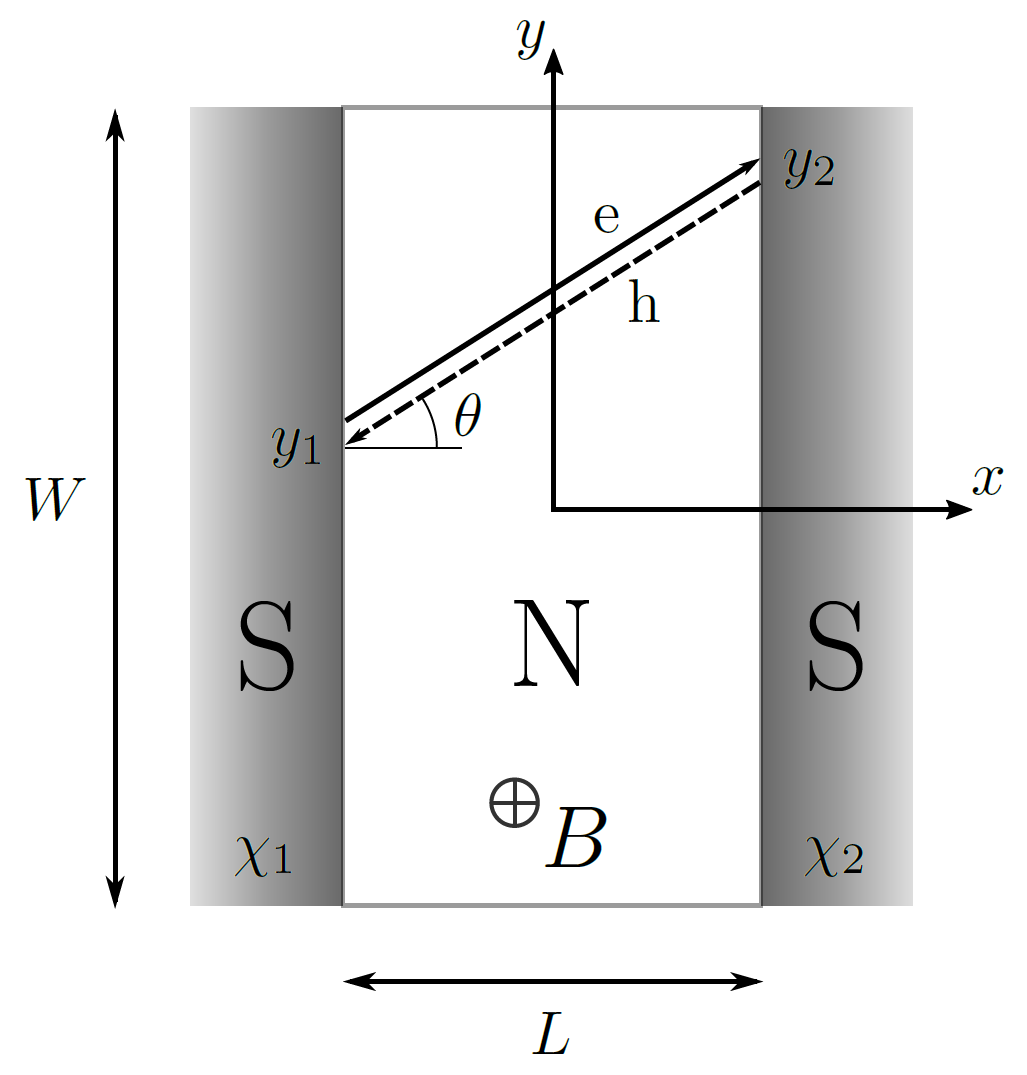}
  \caption{Schematics of the SNS setup at zero split-gate voltage.
    }
  \label{Fig:1}
\end{figure}

\begin{equation}
I_c(\phi)=\text{max}_\chi\{J(\chi,\phi)\}.
\label{critcur}
\end{equation}
Here $\phi=\Phi/\Phi_0=BWL/\Phi_0$ is the dimensionless magnetic flux through the sample in
units of $\Phi_0=\pi \hbar c/e$.

Since the thermal length $L_T\sim \hbar v/(k_B T)$ for the experimental temperature
is much larger than $L$, we set $T=0$.
Following Ref.~\cite{Barzykin1999}, we write the Josephson current for a long junction ($L\gg \xi$) as an integral over the end points
of the Andreev tubes (Fig.~\ref{Fig:1})
\begin{equation}
J(\chi,\phi)=\frac{2e v_F}{\pi \lambda_F W L}
\iint_{-W/2}^{W/2}dy_1 dy_2 \dfrac{\mathcal{J}[\tilde{\chi}(y_1,y_2)]}{\left[1+\left(\frac{y_1-y_2}{L}\right)^2\right]^{3/2}} ,
\label{Jdef}
\end{equation}
where $\mathcal{J}$ is the dimensionless partial Josephson current associated with points $y_1$ and $y_2$
and $\tilde{\chi}(y_1,y_2)$ is the effective phase difference in magnetic field.
Each straight trajectory connecting points $y_1$ and $y_2$ is characterized by angle $\theta$ between the trajectory and $x$-axis,
$
\tan\theta=(y_2-y_1)/L.
$
In the long-junction limit, the partial current is given by
\begin{equation}
\mathcal{J}(\chi)=\sum_{k=1}^\infty \frac{(-1)^{k+1} \mathcal{T}^k}{k} \sin(k \chi)
=\text{Im}\left[ \ln\left(1+\mathcal{T} e^{i \chi}\right)\right], \quad  \xi\ll L,
\label{sawT}
\end{equation}
where we introduced the transmission probability $\mathcal{T}\leq 1$.
For $\mathcal{T}\ll 1$ only the $k=1$ term is important, leading to
\begin{equation}
\mathcal{J}(\chi)\simeq  \mathcal{T} \sin\chi, \quad \mathcal{T}\ll 1,
\label{sawT<1}
\end{equation}
which is the conventional Josephson relation (also valid for $\mathcal{T}\ll 1$ in the short-junction limit).

To include the magnetic field into the consideration, it is convenient
to choose the $x$ dependent gauge for the vector potential as in Ref.~\cite{Meier2016}
(assuming small London penetration length for superconducting leads):
\begin{equation}
\mathbf{A}=A_y \mathbf{e}_y, \quad
A_y=\left\{
                                         \begin{array}{ll}
                                           -B x, & -L/2\leq x\leq L/2, \\[0.2cm]
                                           -\frac{1}{2}B L \sign{x}, &\quad |x|>L/2.
                                         \end{array}
                                       \right.
\label{Ay}
\end{equation}
For such a vector potential, the phase difference due to the magnetic phase acquired on straight trajectories connecting the two interfaces vanishes.
At the same time, the superconducting phases at the interfaces become functions of $y$~\cite{Meier2016}:
\begin{equation}
\label{tildechiAy}
\tilde{\chi}(y_1,y_2)=\chi-\frac{\pi \phi (y_1+y_2)}{W}.
\end{equation}

\begin{figure}
  \centering
  \includegraphics[width=.35\textwidth]{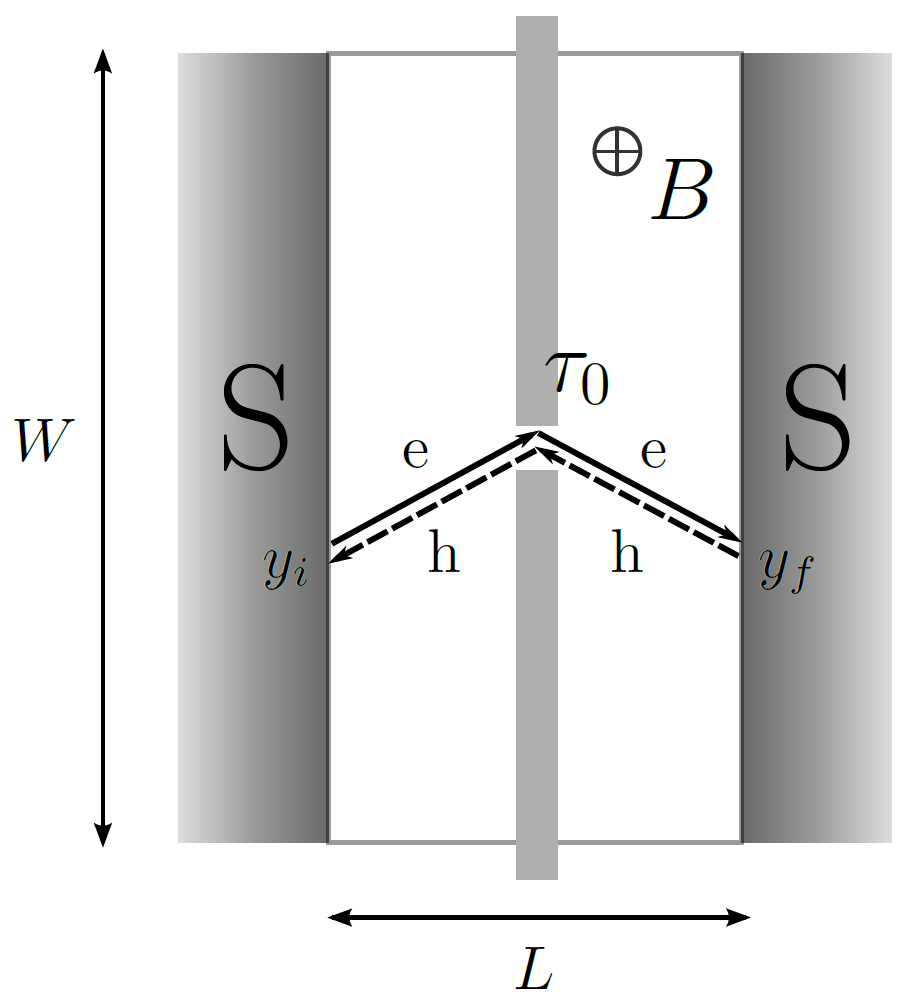}
  \caption{QPC setup with split-gate.
    }
  \label{Fig:6}
\end{figure}

Let us now employ the formalism described above to the QPC setup.
For simplicity, we neglect the geometrical width of the infinitely strong barriers.
We assume that the setup is symmetric, \textit{i.e.}, the QPC is located at $x=0$ and $y=0$.
The QPC has the width which is of the order of (or smaller than) $\lambda_F$ and hence is
approximately characterized by an isotropic transmission probability. 
For low transmission $\mathcal{T}_0\ll 1$, one can retain only the conventional first harmonics in
the partial Josephson currents, Eq.~(\ref{sawT<1}). This implies that under these assumptions, 
the shape of the magnetic interference pattern is, in fact, not sensitive to the relation between $\xi$ and $L$
(long \textit{vs.} short junction).

In terms of the quasiclassical trajectories, the only possible trajectory connecting the
points $y_i$ and $y_f$ at the opposite interfaces should pass through the QPC (here
we discard the boundary scattering). The trajectory is now parameterised by
the two angles: $\theta_i$, corresponding to the velocity in the region $-L/2<x<0$,
and $\theta_f$ in the region $0<x<L/2$ after transmission through the QPC.
These angles satisfy the relations:
\begin{eqnarray}
\tan\theta_i=-\frac{2y_i}{L}, \quad \tan\theta_f=\frac{2y_f}{L}.
\end{eqnarray}

With the gauge (\ref{Ay}), the magnetic phase acquired within the sample reads:
\begin{eqnarray}
\frac{2\pi}{\Phi_0} \int d\mathbf{l} \cdot \mathbf{A}  &=&
-\frac{\pi B}{\Phi_0}\left(\frac{L}{2}\right)^2
\left(-\tan\theta_i + \tan\theta_f\right) =
-\frac{\pi \phi (y_i+y_f)}{2 W}.
\label{phaseQPC}
\end{eqnarray}
The total phase difference is given by the sum of the magnetic phase (\ref{phaseQPC}) 
and the superconducting phase difference in the presence of magnetic field [Eq.~(\ref{tildechiAy})]:
\begin{equation}
\tilde{\chi}(y_i,y_f)=\chi-\frac{3\pi \phi }{2W}(y_i+y_f).
\end{equation}
The total Josephson current now reads
\begin{eqnarray}
J_\text{QPC}(\chi,\phi)
&=&\frac{8e v_F}{\pi \lambda_F W L} \mathcal{T}_0
\sin\chi
\int_{-W/2}^{W/2}dy_i \frac{\cos\frac{3\pi \phi y_i}{2W}}{1+\left(\frac{2y_i}{L}\right)^2}
\int_{-W/2}^{W/2}dy_f
\frac{\cos\frac{3\pi \phi y_f}{2W}}{\left[1+\left(\frac{2y_f}{L}\right)^2\right]^{3/2}}.
\label{JdefQPC}
\end{eqnarray}
The critical current is obtained by the replacement $\sin\chi\to 1$
in the above expression.
At $\phi=0$ we get
\begin{equation}
I_c(0)\equiv I_{c0}=\frac{8e v_F}{\pi \lambda_F} \mathcal{T}_0
\frac{L}{\sqrt{L^2+W^2}}\arctan\frac{W}{L}.
\label{Ic-0}
\end{equation}
The parabolic asymptotics of the critical current at small $\phi$ is found by expanding
the cosine factors in Eq.~(\ref{JdefQPC}):
\begin{eqnarray}
I_c(\phi)&\simeq& I_{c0} \left[1-f_0\left(\frac{W}{L}\right)\phi^2\right],\qquad \phi\to 0,\\
f_0(x)&=&\frac{9\pi^2}{32x^2} \left(\frac{x}{\arctan x}+\frac{\sqrt{1+x^2}}{x}\ln(x+\sqrt{1+x^2})-2\right).
\label{small-phi}
\end{eqnarray}
In the opposite limit of high fields, $\phi\to \infty$, we extend the integration over $y_i$ and $y_f$
to $\pm \infty$ and obtain
\begin{eqnarray}
I_c(\phi)&\simeq& I_{c0}\ \frac{\pi^{3/2}}{4} \frac{\sqrt{W^2+L^2}}{W\arctan(W/L)}
\sqrt{\frac{3\pi L\phi}{2W}}
\exp\left(-\frac{3\pi L}{2W}\phi\right), \qquad \phi\to\infty.
\label{large-phi}
\end{eqnarray}
The evaluation of the integrals in Eq.~(\ref{JdefQPC}) in the whole range of magnetic fields yields the curve 
for the normalized critical current shown in Fig. 5b of the main text. The width of the analytical curve was fitted to the experimental
one thus accounting for the actual geometry and those factors (\textit{e.g.}, the finite width of the barriers and the finite magnetic penetration depth)
that were neglected in the simplified model.

\subsection{Numerical model: Short junction}

\subsubsection{Definition of the studied system}

\begin{figure*}
\includegraphics[width=0.3\textwidth]{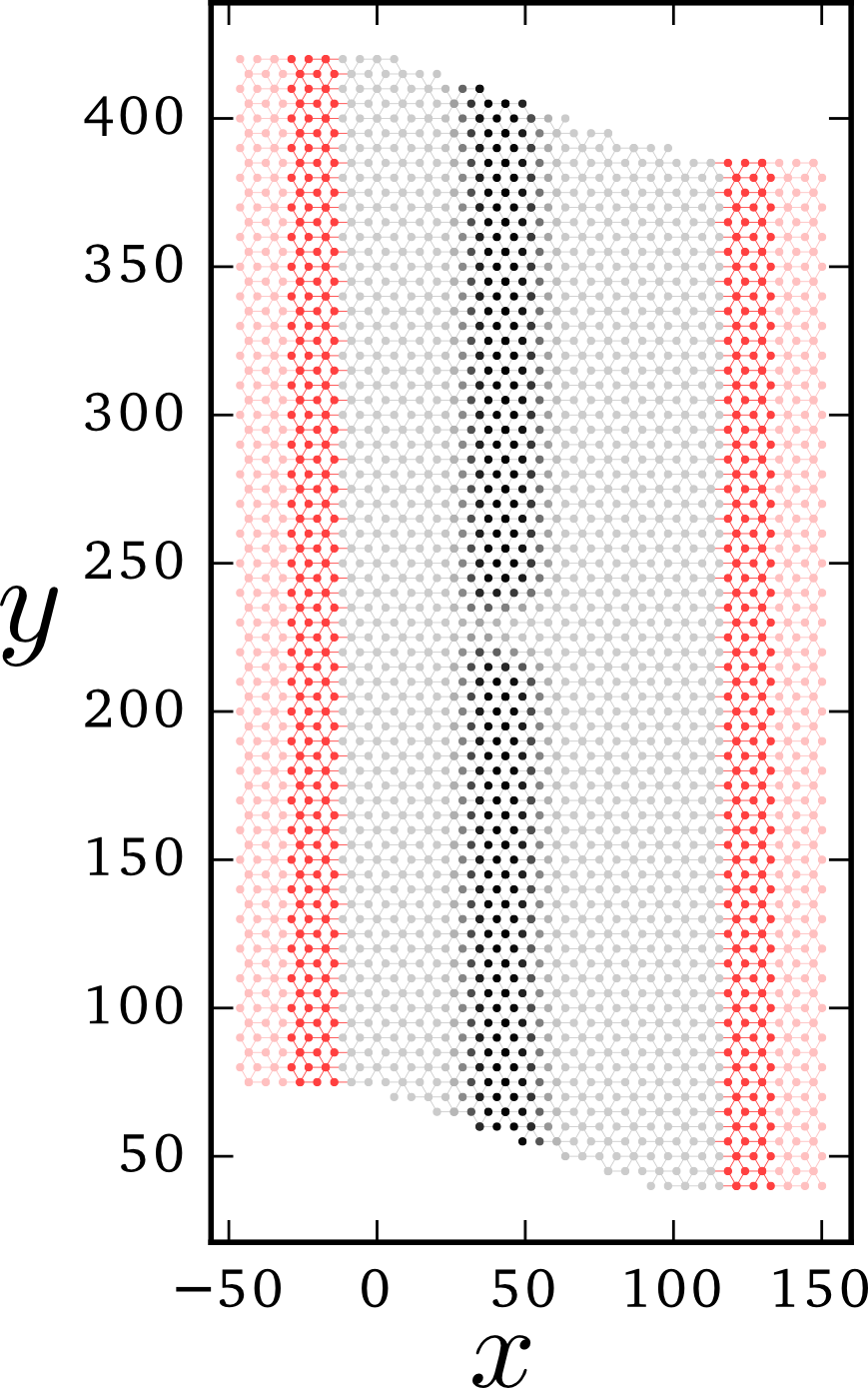}
\caption{
\label{Schematic}
Illustration of the setup used for the simulations. 
The darker region represents the QPC area while the red strips represent the leads attached to the scattering region. 
The scales of the $x$- and $y$-axis are defined in terms of the tight-binding lattice constant.}
\end{figure*}

We study a supercurrent through a superconductor--bilayer-graphene--superconductor (SBLGS) Josephson junction (JJ) 
in the presence of top split gates in the experimental geometry, as shown in Fig.~\ref{Schematic}. 
In particular, we investigate the orbital effects of the magnetic field which is applied perpendicular 
(say in $z$-direction) to the junction in the normal region. 
For numerical simulations, we use a Landau gauge such that the vector potential is given by $\mathbf{A} = (-By \hat{x}, 0, 0)$.

\subsubsection{Short-junction supercurrent calculation} \label{short-JJ}

Here, we explain the formalism used for calculating the supercurrent numerically. 
We restrict ourselves to the short-junction limit and use the scattering matrix approach. 
For transport below the superconducting energy gap, an electron incident on the superconducting interface 
is Andreev reflected as a hole, which results in emergence of Andreev bound states. 
To calculate the spectrum of these bound states, we use the two types of scattering matrices. 
The matrix $s_N$, which is due to the normal region and is block diagonal in electron and hole space, is defined as
\begin{equation}
s_N(\epsilon) = \left[ {\begin{array}{cc}
s(\epsilon) & 0 \\
0 & s^*(-\epsilon) \\
\end{array} } \right],
\end{equation}
while the other matrix $s_A$ is due to Andreev reflections and is block off-diagonal as it couples electrons with holes:
\begin{equation}
s_A(\epsilon) = \alpha (\epsilon)\left[ {\begin{array}{cc}
0 & r_{A}^* \\
r_{A} & 0 \\
\end{array} } \right],
\end{equation}.
with 
$$\alpha(\epsilon) = \exp[-i \arccos(\epsilon/\Delta)]$$ 
and $r_A$ a diagonal matrix
\begin{equation}
r_A = \left[ {\begin{array}{cc}
i e^{i\chi/2} \mathbf{1}_{n_1} & 0 \\
0 & i e^{-i \chi/2} \mathbf{1}_{n_2} \\
\end{array} } \right].
\end{equation}

In general, these two matrices depend on the energy $\epsilon$ of the state. 
However, in the short-junction limit, we have $s(\epsilon) \approx s(-\epsilon) \approx s(0) \equiv s$. 
The condition on these scattering matrices to have a bound state is given in Ref.~\cite{Beenakker1991}:
\begin{equation}
s_{A}(\epsilon)s_{N}(\epsilon)\Psi_{in} = \Psi_{in},
\end{equation}
where $\Psi_{in} = (\Psi_{in}^e, \Psi_{in}^h)$ is a vector of complex coefficients describing a wave incident on the junction in the basis of modes incoming from the superconducting leads into the normal region. By using the individual matrices, we arrive at the following eigenvalue problem:
\begin{equation}
\left[ {\begin{array}{cc}
s^{\dagger} & 0 \\
0 & s^{T} \\
\end{array} } \right] \left[ {\begin{array}{cc}
0 & r_{A}^* \\
r_{A} & 0 \\
\end{array} } \right]\Psi_{in}=\alpha \Psi_{in}.
\end{equation}
By applying Joukowsky transform, we can map the above eigenvalue problem from  $\alpha$ to $\epsilon/\Delta$ \cite{vanHeck2014}:
\begin{equation}
\left[ {\begin{array}{cc}
0 & -iA^{\dagger} \\
iA & 0 \\
\end{array} } \right]\Psi_{in}=\frac{\epsilon}{\Delta} \Psi_{in},
\end{equation}
where 
$$A \equiv \frac{1}{2}(r_A s -s^T r_A).$$ 
For only electron states, we can write from the above equation
\begin{equation}
A^{\dagger}A \Psi_{in}^e = \frac{\epsilon^2}{\Delta^2} \Psi_{in}^e.
\end{equation}

In the limit of first order perturbation theory, we can write an expression for the differential of bound state energy with respect to the superconducting phase as
\begin{equation}
\frac{d\epsilon}{d\chi}=\frac{\Delta^2}{2\epsilon} \frac{\left\langle \Psi_{in} \right| \frac{d(A^{\dagger}A)}{d\chi}\left|\Psi_{in}\right\rangle}{\left\langle \Psi_{in} \right|\left.\Psi_{in}\right\rangle}.
\end{equation}
The advantage of having the above expression is that one do not need to take a numerical differentiation of bound state energy over $\chi$. Instead we just solve an eigenvalue problem for this differential. As a result, we obtain the Josephson current as
\begin{equation}
J(\chi) = -\frac{2e}{\hbar}\sum_p \tanh(\epsilon_p/2k_BT)\frac{d\epsilon_p}{d\chi}.
\end{equation}
The critical current $I_c$ is then obtained from Eq.~(\ref{critcur}).

\subsubsection{Tight-Binding / Schr\"{o}dinger equation}

The tight-binding Hamiltonian for the normal scattering region can be written as
\begin{equation}
\label{eq:hamiltonian}
H =-t \sum_{i, j, m} e^{\iota \phi_{ij}} a_{mi}^{\dagger} b_{mj} - \gamma_{1} \sum_j a_{1j}^{\dagger} b_{2j} - \sum_{i, m} (\mu_{i} - (-1)^m \delta_{i}) (a_{mi}^{\dagger}a_{mi} + b_{mi}^{\dagger}b_{mi}) +H.c.,
\end{equation}
with  the magnetic phases $\phi_{i,j} = \frac{2\pi e}{h} \int_{i}^{j} \mathbf{A} \cdot d\mathbf{r}$. 
The indices $i, j$ corresponds to the lattice sites, whereas $m = 1, 2$ denote the two layers of the BLG. 
The operators $a_{mi}, a_{mi}^{\dagger}$ ($b_{mi}, b_{mi}^{\dagger}$) are the annihilation and creation operators for electrons at site $i$ in sublattice $A_m$ ($B_m$), respectively. The parameters $t$ and $\gamma_1$ are the intralayer and interlayer (between dimer sites) hopping constants, respectively, whereas $\mu$ and $\delta$ corresponds to the onsite energies given by
\begin{equation}
\mu_{i}=(\varphi_{BG}+u_i \varphi_{SG})/2,
\end{equation}
and
\begin{equation}
\delta_{i} = -(u_i \varphi_{SG}-\varphi_{BG})/\eta.
\end{equation}
Here $u_i$ defines the gated region such that
\begin{equation}
u_i = \begin{cases}
    1 & \text{if $i$ is in gated region}\\
    0 & \text{if $i$ is outside gated region}
  \end{cases}
\end{equation}
and $\eta$ is the numerical factor accounting for the geometry of the setup in the direction perpendicular
to the graphene plane. In what follows, we choose $\eta=2.5$.

In the above equations, $\varphi_{BG}$ and $\varphi_{SG}$ represent the strengths of the on-site potentials 
introduced by the back gate and top split gate, respectively.
The corresponding Bogoliubov-De Gennes Hamiltonian for the SBLGS junction reads:
\begin{equation}
H_{BdG} = \begin{pmatrix}H & \Delta \\ \Delta^* & -H^*\end{pmatrix},
\end{equation}
where $\Delta$ is a step-like function given by
\begin{equation}
\Delta = \begin{cases}
    \Delta_0 e^{i\chi_1} & \text{left lead}\\
    \Delta_0 e^{i\chi_2} & \text{right lead} \\
    0 & \text{scattering region}
  \end{cases}.
\end{equation}

\begin{figure*}
\includegraphics[width=1\textwidth]{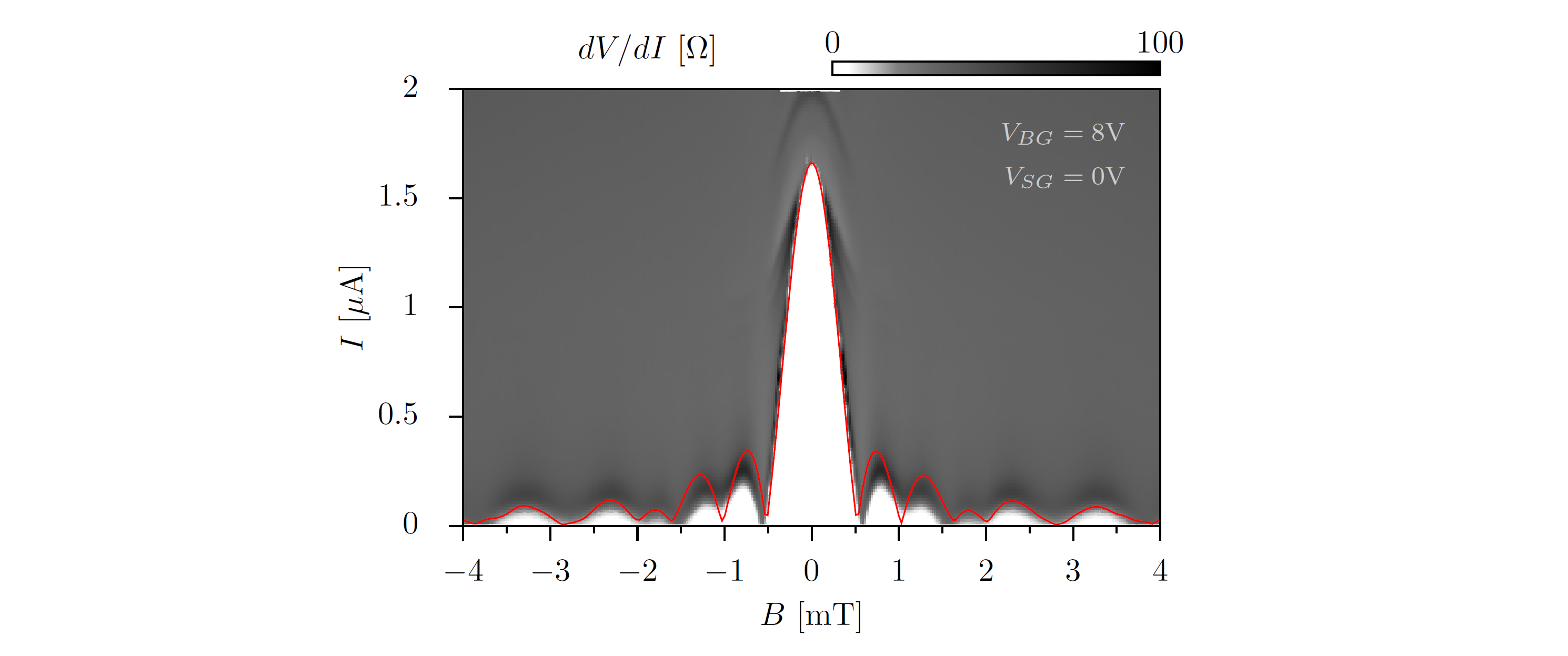}
\caption{
\label{NumericsFP}
Fit (red curve) of experimental magnetic interference pattern $dV/dI(B,I)$ at zero split-gate voltage (gray-scale map).}
\end{figure*}

To calculate the scattering matrix $s_N$ introduced in the previous section, we use Kwant \cite{Groth2014} where we discretise Eq.~(\ref{eq:hamiltonian})  on a honeycomb lattice. 
To verify our model, we fit the experimental data at zero split-gate voltage Fig.\,\ref{NumericsFP}. The corresponding magnetic interference pattern appears quite distorted compared to a $sinc$ function of a regular, short and wide rectangular junction \cite{tinkhambook}. The fitting is done by rescaling both axes to match the experimental plot using a junction area of $\sim 4.60 \cdot 10^{-12}\,$m$^2$ (corresponding to $\lambda_L=245$\,nm, close to the value extracted from the analytical model $\lambda_L=275$\,nm). It shows that the missing/suppressed lobs of the experimental magnetic interference pattern at zero split-gate voltage match the ones obtained from simulation at some finite strength 
of the potential in the gated area ($\varphi_{SG}/t = -0.024$), which may indicate the presence of a small but non-zero potential in the gated region in our experiments. This potential can result from the charge redistribution in the metallic top-gate which is induced by the electrostatic interaction between electrons in the gate and in bilayer graphene.


\subsubsection{Parameters used for the simulation of $I_c(B,\varphi_{SG})$}

Here, we describe the parameters used in the simulations displayed in Fig.\,5c of the main text.
We take $\varphi_{BG} = 0.2 t$, $\gamma_1 = 0.4 t$ and vary $\varphi_{SG}$ starting from the ``open'' regime (ungapped BLG in gated region) to the ``closed'' regime (gapped BLG in the gated region). The numerical calculations are performed in the short-junction limit and at finite temperature $k_BT = \Delta_0/4$. 
This value of $T$, although being higher than
the actual experimental temperature, corresponds to the thermal length that is much larger than $L$. Therefore, the difference is negligible; at the same time,
the higher $T$ used in the simulation substantially simplifies the numerics.
